\documentclass[twocolumn]{aastex631}

\usepackage{soul}
\usepackage{xcolor}

\begin{document}

%\title{Hidden Companions: Tracking Down Main Sequence Partners in White Dwarfs with Gaia XP Spectra and Self-organizing Maps}
\title{Finding White Dwarfs' Hidden Companions using an Unsupervised Machine Learning Technique}

\author[0000-0001-5797-252X]{Xabier P\'erez-Couto}
\affiliation{Universidade da Coruña (UDC), Department of Computer Science and Information Technologies,\\ Campus de Elviña s/n, 15071, A Coruña, Galiza, Spain}
\affiliation{CIGUS CITIC, Centre for Information and Communications Technologies Research,\\ Universidade da Coruña, Campus de Elviña s/n, 15071 A Coruña, Galiza, Spain}

% \author[0000-0001-9296-3100]{Lara Pallas-Quintela}
% \affiliation{Universidade da Coruña (UDC), Department of Computer Science and Information Technologies,\\ Campus de Elviña s/n, 15071, A Coruña, Galiza, Spain}
% \affiliation{CIGUS CITIC, Centre for Information and Communications Technologies Research,\\ Universidade da Coruña, Campus de Elviña s/n, 15071 A Coruña, Galiza, Spain}

\author[0000-0002-7711-5581]{Minia Manteiga}
\affiliation{Universidade da Coruña (UDC), Department of Nautical Sciences and Marine Engineering,\\ Paseo de Ronda 51, 15011, A Coruña, Galiza, Spain}
\affiliation{CIGUS CITIC, Centre for Information and Communications Technologies Research,\\ Universidade da Coruña, Campus de Elviña s/n, 15071 A Coruña, Galiza, Spain}
\affiliation{AIRExS, CITIC/UDC, Unidad Asociada al Instituto de Astrofísica de Andalucía, CSIC}

\author[0000-0003-4936-9418]{Eva Villaver}
\affiliation{Instituto de Astrofísica de Canarias, 38200 La Laguna, Tenerife, Spain}
\affiliation{Universidad de La Laguna (ULL), Astrophysics Department,\\ 38206 La Laguna, Tenerife, Spain}

% \author[0000-0003-4693-7555]{Carlos Dafonte}
% \affiliation{Universidade da Coruña (UDC), Department of Computer Science and Information Technologies,\\ Campus de Elviña s/n, 15071, A Coruña, Galiza, Spain}
% \affiliation{CIGUS CITIC, Centre for Information and Communications Technologies Research,\\ Universidade da Coruña, Campus de Elviña s/n, 15071 A Coruña, Galiza, Spain}

%% Note that the \and command from previous versions of AASTeX is now
%% depreciated in this version as it is no longer necessary. AASTeX 
%% automatically takes care of all commas and "and"s between authors names.

%% AASTeX 6.31 has the new \collaboration and \nocollaboration commands to
%% provide the collaboration status of a group of authors. These commands 
%% can be used either before or after the list of corresponding authors. The
%% argument for \collaboration is the collaboration identifier. Authors are
%% encouraged to surround collaboration identifiers with ()s. The 
%% \nocollaboration command takes no argument and exists to indicate that
%% the nearby authors are not part of surrounding collaborations.

\received{}
\accepted{}

%% Mark off the abstract in the ``abstract'' environment. 
\begin{abstract}

White dwarfs (WD) with main-sequence (MS) companions are crucial probes of stellar evolution. However, due to the significant difference in their luminosities, the WD is often outshined by the MS star. The aim of this work is to find hidden companions in Gaia's sample of WD candidates. Our methodology involves applying an unsupervised machine learning algorithm for dimensionality reduction and clustering, known as Self-Organizing Map (SOM), to Gaia BP/RP (XP) spectra. This strategy allows us to naturally separate WDMS binaries from single WDs from the detection of subtle red flux excesses in the XP spectra that are indicative of low-mass MS companions. We validate our approach using confirmed WDMS binaries from the SDSS and LAMOST surveys, achieving a precision of $\sim 90\%$. We demonstrated that the luminosity of the faint companions in the missed systems is $\sim 50$ times lower than that of their WD primaries.
Applying our SOM to 90,667 sources, we identify 993 WDMS candidates, 506 of which have not been previously reported in the literature. If confirmed, our sample will increase the known WDMS binaries by $20\%$. Additionally, we use the Virtual Observatory Spectral Energy Distribution Analyzer (VOSA) tool to refine and parameterize a ``golden sample'' of 136 WDMS binaries through multi-wavelength photometry and a two-body Spectral Energy Distribution fitting. These high-confidence WDMS binaries are composed by low-mass WDs ($\sim 0.42 M_{\odot}$), with cool MS companions ($\sim 2800$ K). Finally, 13 systems exhibit periodic variability consistent with eclipsing binaries, making them prime targets for further follow-up observations.

\end{abstract}
%% Keywords should appear after the \end{abstract} command. 
%% The AAS Journals now uses Unified Astronomy Thesaurus concepts:
%% https://astrothesaurus.org
%% You will be asked to selected these concepts during the submission process
%% but this old "keyword" functionality is maintained in case authors want
%% to include these concepts in their preprints.

%% From the front matter, we move on to the body of the paper.
%% Sections are demarcated by \section and \subsection, respectively.
%% Observe the use of the LaTeX \label
%% command after the \subsection to give a symbolic KEY to the
%% subsection for cross-referencing in a \ref command.
%% You can use LaTeX's \ref and \label commands to keep track of
%% cross-references to sections, equations, tables, and figures.
%% That way, if you change the order of any elements, LaTeX will
%% automatically renumber them.
%%
%% We recommend that authors also use the natbib \citep
%% and \citet commands to identify citations.  The citations are
%% tied to the reference list via symbolic KEYs. The KEY corresponds
%% to the KEY in the \bibitem in the reference list below. 

\section{Introduction} \label{sec:intro}

It is well established that the binary fraction of stars is highly dependent on the stellar mass, ranging from 30\% for M-type stars \citep{wintersetal2019} to 70\% for O and B-type stars \citep{sanaetal2014, moeanddistefano2017}, with a mean incidence of 50\%  for solar-type stars \citep{raghavanetal2010}. 

The more massive star in the pair will evolve faster and, if it is a low-to-intermediate-mass star ($\lesssim 8$ $\mathcal{M}_{\odot}$), it will eventually become a white dwarf (WD, \citet{ibenetal1997}) forming a WD plus main-sequence (MS) star binary (hereafter, WDMS). Given the very predictable cooling age of the WDs, WDMS binary pairs are excellent cosmic clocks that have been used to study fundamental astrophysical parameterizations such as the age–metallicity relation \citep{mansergasetal2021a}, the initial -to -final mass  \citep{zhaoetal2012}, and the mass-radius relation \citep{raddietal2025}.

Different outcomes are expected for the WDMS binary depending on the orbital separation. In wide orbits pairs, the MS companion evolves independently eventually leading to the formation of a WD--WD binary. Conversely, close WDMSs are susceptible to undergo mass transfer episodes, potentially leading to Cataclysmic Variables \citep[CVs;][]{parsonsetal2013, sunetal2021}, Novae, Symbiotic, and Type Ia Supernovae (SNe) \citep{wangandhang2012},  essential tools in cosmological and stellar evolution studies \citep{leibundgutandsullivan2018}.

The most extensive samples of WDMS to date are those obtained by the Sloan Digital Sky Survey \citep[SDSS, see e.g.][]{mansergasetal2016} and the Large Sky Area Multi-Object Fiber Spectroscopic Telescope \citep[LAMOST, see e.g.][]{renetal2018} with a total of 4100 WDMSs. However, both surveys exhibit certain observational biases against cool WDMS, resulting in an apparent absence of systems with $T_{eff} < 10,000$K.

Several studies have demonstrated the feasibility of automatically identifying WDMS binaries using machine learning techniques with promising accuracy \citep[around 80$\%$ using Random Forest; see][]{echeverryetal2022}, and successfully detecting candidates in open clusters with Support Vector Machines \citep[SVM;][]{grondinetal2024}. \citet{kaoetal2024} in particular, identified an isolated group of 1,096 WDMS candidates by using a Uniform Manifold Approximation and Projection (UMAP) through the largest white dwarf catalog available to date. 
Recently, we have used Self-Organizing Maps \citep[SOMs;][]{kohonen1982}, an unsupervised neural network-based algorithm to find polluted WD candidates based on Gaia XP spectra \citep{perezcoutoetal2024}. 

In this work, we will use a similar methodology to that used in \citet{perezcoutoetal2024} to identify MS companions in WD spectra from the catalog of \citet{gentilefusilloetal2021}. This catalog is built using color-magnitude and astrometric cuts to prioritize single WDs. Therefore, any secondary companion to a WD in the sample is expected to be a low-mass, late-type M dwarf or even a brown dwarf, as its presence is not expected to significantly affect the photometry or astrometry of the WD.

The paper is organized as follows: in Section 2, we describe the data used and the SOM learning process, in \S3 we apply the method to the data and discuss the results. Finally, in \S4 we summarize our main findings and present the conclusions of the paper.

\section{Methodology} \label{sec:data}

The Gaia Mission \citep{gaiacollaboration2023} has provided, in its Third Data Release (DR3), high-quality astrometric data and photometry from the Blue (BP) and Red Photometers (RP) for 1460 million sources of our Galaxy. This extensive dataset has been instrumental in identifying new WDMS by using the Gaia $G$, $G_{BP}$, and $G_{RP}$ Color Magnitude Diagram (CMD) and Virtual Observatory (VO) tools. In particular, the Virtual Observatory Spectral Energy Distribution Analyzer \citep[VOSA\footnote{
\url{http://svo2.cab.inta-csic.es/theory/vosa/}},][]{bayoetal2008} allowed \cite{mansergasetal2021b} to find 97 new WDMS and parameterize their stellar properties. 

In addition to the BP/RP photometry, Gaia published low-resolution ($R \approx 70$) BP/RP spectra (hereafter, XP spectra) for about 220 million sources \citep{deangelietal2023}. Instead of flux units per wavelength unit, each XP spectrum is given as an array of 110 coefficients of a series of Hermite basis functions (55 for BP and 55 for RP). Given the infeasibility of visually inspecting such an extensive data set, numerous studies have employed machine learning (ML) algorithms to mine the data in the search and classification of WD \citep{garciazamoraetal2023, vincentetal2024, kaoetal2024, perezcoutoetal2024, 2025arXiv250505560G}. 

Self-Organizing Maps \citep[SOMs;][]{kohonen1982}, is an unsupervised neural network-based algorithm that combines either dimensionality reduction ---to project the XP coefficients in a two-dimensional grid map--- and cluster ---to group similar elements together in the same neuron---. The power of this dual technique demonstrates that SOMs are a useful artificial intelligence tool for object classification in various fields of astrophysics\citep[see e.g.][]{torresetal1998, naimetal2009, ordonezetal2010, geach2012, wayandklose2012, fustesetal2013a, fustesetal2013b, carrascoandbrunner2014, dafonteetal2018, alvarezetal2022, perezcoutoetal2024}.

\subsection{Input data} \label{subsec:initialsample}

The initial sample is based on the \citet{gentilefusilloetal2021} catalog, where a large sample of WD candidates is selected first by imposing the following cut in the Gaia CMD:
\begin{equation}\label{WDlocus}
    G_{abs} > 6 + 5 \times \left(G_{BP} - G_{RP}\right),
\end{equation}
a \texttt{parallax\_over\_error > 1} and several additional quality cuts to discard bad astrometric solutions up to a final sample size of $1.3$ million sources. 

This color cut is indeed not the most effective way of identifying a large number of WDMS binaries, as it excludes WDs situated in the CMD between the WD locus and the MS branch ---a region above which approximately $90\%$  of WDMS binaries are expected to be found, according to recent population synthesis simulations \citep{mansergasetal2021b, santosgarciaetal2025}---. Nevertheless, we adopt this color cut in the present study, which specifically focuses on the WD region. This approach ensures that any detected companion has low emission, as the WD dominates, making very low-mass companions, such as M stars or brown dwarfs, the most likely candidates.

Some astrometric cuts used in the \citet{gentilefusilloetal2021} catalog such as the Renormalized Unit Weight Error (RUWE) $< 1.1$, \texttt{ipd\_gof\_harmonic\_amplitude < 1}, or \texttt{astrometric\_excess\_noise\_sig <2} efficiently clean the sample from the majority of astrometric contaminants (among them, many unresolved binaries) \citep{belokurovetal2020}. This, in conjunction with the fact that they are unresolved despite their proximity, makes any WDMS binary found in their catalog to be a very close binary.

In \citet{gentilefusilloetal2021}, the authors computed a probability of an object being a WD ($P_{WD}$). This probability is determined using a reference dataset of $22,998$ spectroscopically confirmed WDs and $7124$ contaminants identified through visual inspection in the SDSS. These datasets are modeled as normalized 2D Gaussian distributions, producing distinct density maps for WDs and contaminants. The $P_{WD}$ for each candidate is calculated by integrating its CMD Gaussian representation with a map formed by taking the ratio of the WD density map to the combined density of both WDs and contaminants.

\begin{figure*}[ht]
\gridline{\fig{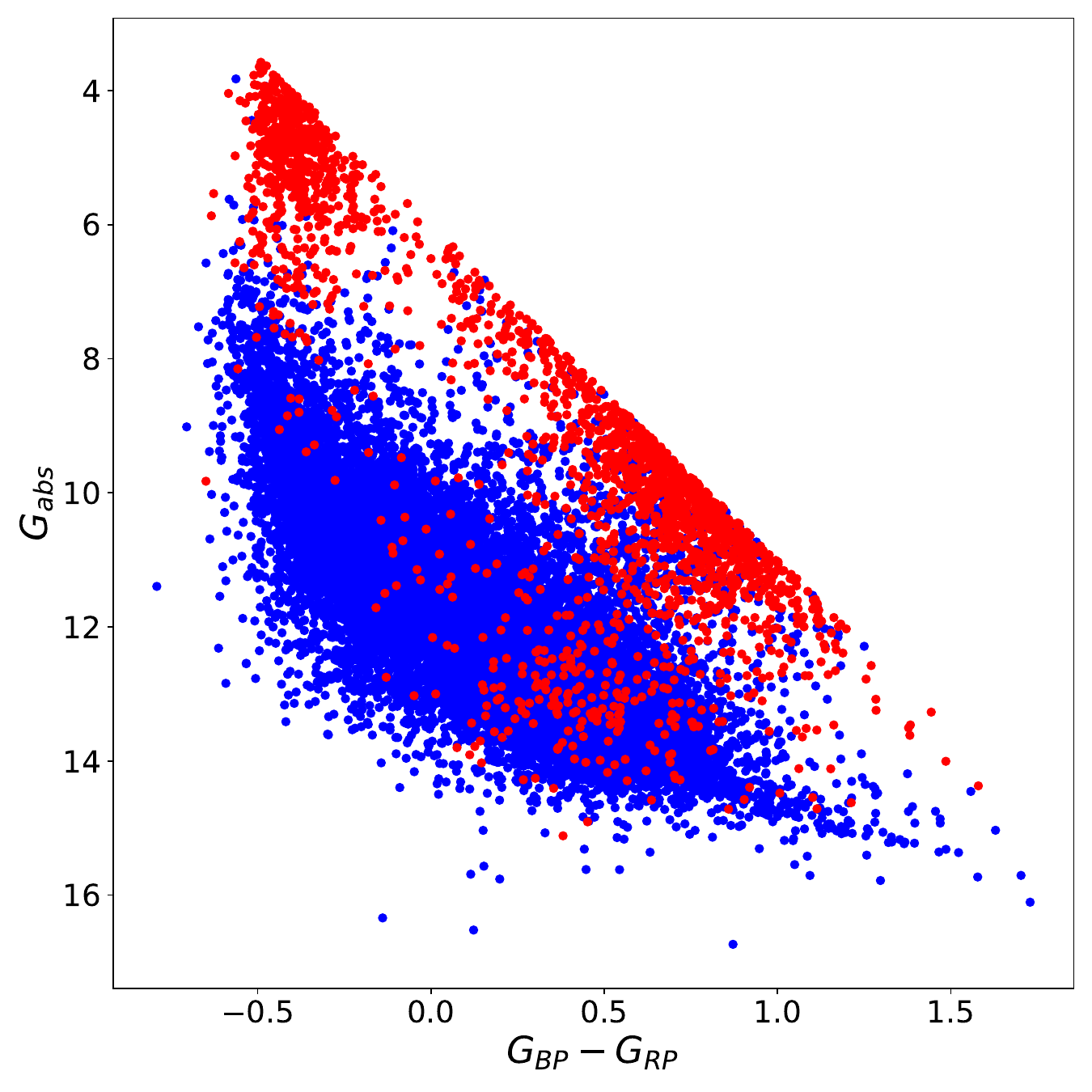}{0.48\textwidth}{(a) $\bar{\omega}/\sigma_{\bar{\omega}}>1$} \label{parallax_over_error>1}
          \fig{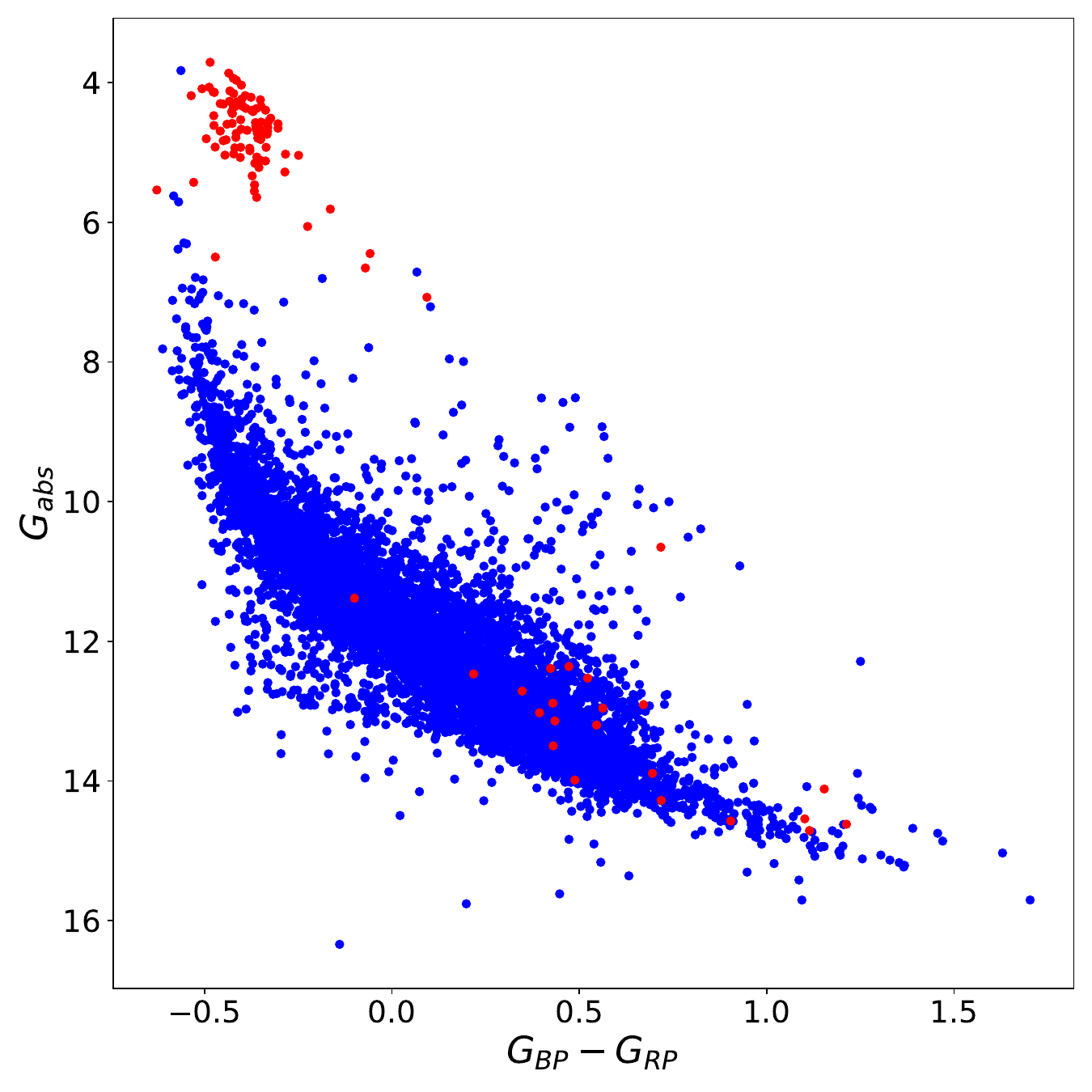}{0.48\textwidth}{(b) $\bar{\omega}/\sigma_{\bar{\omega}}>10$} \label{parallax_over_error>10}
}

\caption{\label{SDSSconfirmedVScontaminants} SDSS confirmed vs contaminants CMD with $\bar{\omega}/\sigma_{\bar{\omega}}>1$ (left) and $\bar{\omega}/\sigma_{\bar{\omega}}>10$ (right).}
\end{figure*}

The definition of contaminant used in \citet{gentilefusilloetal2021} included WDMS binaries, and hence a probability filter of, for instance, $P_{WD} > 0.9$, would exclude the majority of contaminants such as QSOs or Galaxies, but also most of the WDMS we aim to discover.
For this reason, we will not use the $P_{WD}$ in the following.

In contrast, we only consider as contaminants those sources with the SDSS spectral class ``QSO'', ``GALAXY'', and ``STAR''. The ``Unreli'' (for unreliable) and ``UNKN'' (for unknown) sources in the Gaia-SDSS sample of \citet{gentilefusilloetal2021} were discarded from the sample since we are not confident to confirm if they are WDs or contaminants. This leaves us with $26,423$ SDSS confirmed WDs (either single or binary sources) and 4588 contaminants.

Subsequently, we use a parallax ($\bar{\omega}$) over error ($\sigma_{\bar{\omega}}$) (or $\bar{\omega}/\sigma_{\bar{\omega}}$) $>10$ that will ensure a more precise $G_{abs}$, and therefore a more reliable location in the CMD. Additionally, we have included these additional filters to ensure the quality of XP spectra: 

i) \texttt{visibility\_periods\_used > 10}, where each visibility period is a group of observations separated from the next by at least 4 days, so that only those sources that were astrometrically well observed are retained \citep{lindegrenetal2018}.

ii) \texttt{(phot\_bp\_n\_obs > 10) \& (phot\_rp\_n\_obs > 10)}, refer to the minimum number of CCD transits for BP and RP spectra, respectively, following the recommendations set forth by \citet{andraeetal2023} to ensure an adequate signal-to-noise ratio (S/N) for subsequent spectral analysis.

iii) \texttt{|phot\_bp\_rp\_excess\_factor\_corrected| < 5 x sigma\_excess\_factor} ensures that the photometry of $G_{BP}$, $G_{RP}$, and $G$ is consistent and free from contamination from external sources in the same field of view, as elucidated by \citet{rielloetal2021}

We obtained the Gaia XP spectra for this sample using the DataLink Gaia tool (available at \url{https://www.cosmos.esa.int/web/gaia-users/archive/datalink-products}) through the \texttt{astroquery} Python package \citep{ginsburgetal2019}.

Finally, a S/N $> 10$ filter was applied through the coefficients. The S/N for both BP and RP spectra was calculated by taking the ratio between the $\mathcal{L}_2$ norm of the BP (RP) array of coefficients and the $\mathcal{L}_2$ norm of the array of BP (RP) coefficient uncertainties. As a result, we obtained an initial sample for our study comprising a total of $90,667$ sources. 

To roughly estimate the contaminant ratio in our sample, as well as the effectiveness of the $\bar{\omega}/\sigma_{\bar{\omega}}$ filter in discarding them, we show in Figure \ref{SDSSconfirmedVScontaminants}b the Gaia CMD with the SDSS confirmed WDs and contaminants that meet the above filters in blue and red, respectively. However, in the CMD of the left (Figure \ref{SDSSconfirmedVScontaminants}a) we relax the parallax-over-error filter up to the original value in \citet{gentilefusilloetal2021}: $\bar{\omega}/\sigma_{\bar{\omega}}>1$, while in the right (Figure \ref{SDSSconfirmedVScontaminants}b) we show the resulting CMD for $\bar{\omega}/\sigma_{\bar{\omega}}>10$. 

As illustrated in Figure \ref{SDSSconfirmedVScontaminants}, the image on the right is visibly more pristine and devoid of contaminants. Indeed, the contaminant fraction has been reduced from 7.5\% (2074 contaminants) to 0.9\% (112 contaminants), indicating that the input sample of the SOM is unlikely to contain a contamination level greater than 1\%.

\subsubsection{Reference  catalogs}\label{subsec:WDMS_catalogs}

Despite the unsupervised nature of the classification process, which does not rely on a training dataset, spectroscopically confirmed WDMS spectra are required as a reference to label the final clusters. As a baseline, we rely on the Montreal White Dwarf Database\footnote{\url{https://montrealwhitedwarfdatabase.com}} (MWDD), which is so far the most complete catalog of WDs based on more than 200 references from the literature \citep{dufouretal2016}, containing information about each WD such as its spectral type or binarity. As of January 30th, 2025, it contains information for $144,800$ WDs. Most of them also belong to the catalog of \citet{vincentetal2024} which is an automatic classification of Gaia DR3 XP spectra based on gradient-boosted decision trees. Despite the great performance shown by their method, their classification is still based on Gaia low-resolution spectra, and thus it is a catalog of WD candidates instead of confirmed WDs.

Therefore, we decided to ignore sources with only low-resolution spectra in order to keep our reference sample of confirmed WDs as clean as possible. To this end, we discarded those sources included in the \citet{vincentetal2024} catalog if they have only one available optical spectrum. For the rest of the MWDD we used all the sources with at least one available spectrum and a confirmed spectral type. We did not include sources with subdwarf (sdO, sdB, \dots) spectra.

From this set, we selected as WDMS sources those with the ``WDMS'' binarity flag, resulting in $2849$ sources. We also included some sources from the MWDD without a positive binarity flag but with a spectral type containing one of the following strings: `+M', `+dM', `+K', `+G', or `+F', which indicate the presence of an MS companion in the source's spectra. This resulted in an updated WDMS count of $3246$, of which $377$ have XP spectra available and that passed the filters described in Section  \ref{subsec:initialsample}.

We also cross-matched this MWDD WDMS sample with the largest WDMS catalogs up to date: the SDSS DR12 WDMS spectroscopic catalog \citep{mansergasetal2010, mansergasetal2012, mansergasetal2013, mansergasetal2016} and the LAMOST DR5 catalog \citep{renetal2014, renetal2018}. As a result, 14 SDSS WDMS that are not classified by the MWDD as binaries have been added, as well as 19 LAMOST WDMS. This resulted in a final WDMS sample of 406 sources (4 sources were duplicated) that will be used as a reference in the labeling process of the SOM. 

The remaining MWDD sources that are not included in the WDMS binary sample and that do not correspond to any other type of binarity (i.e., those with an empty binarity field in the MWDD, and a spectral type without a `+' sign)  are designated as single WD sources ($13,426$ sources), and the remaining sources in our initial sample ($76,835$ sources) are considered candidates in the following.

\subsection{Self-Organizing Maps}\label{som_description}

While most unsupervised machine learning techniques are either utilized for dimensionality reduction (e.g. t-SNE, UMAP) or clustering (e.g. K-means, DBSCAN), SOMs integrate both applications within a single neural network-based algorithm. Indeed, given a high-dimensional nonlinear data set (in our case, constructed from arrays of 110 coefficients per spectrum), the SOM projects each element on a two-dimensional map, where analogous elements are assigned to the same neuron. Moreover, neurons with similar subpopulations are also grouped in the map, while very different subpopulations are highly distanced. This results in preserving the topology order, allowing for the recognition of patterns in the data. Furthermore, the clustering of neurons into closed groups allows the accurate delineation and classification of these populations.

Once the dimensions of the map, $M \times N$ have been established, the learning process starts with a random initialization of the weight, $\mathbf{w}_{m,n}$, of each neuron, $\mathbf{z}_{m,n}$. Each $\mathbf{w}_{m,n}$ is a random array of 110 elements. After that, the first iteration takes each XP spectrum, $\mathbf{x_i}$, and looks for the winner neuron or Best Matching Unit (BMU), $\mathbf{z}_{m,n}$, by minimizing the distance (for example, the Euclidean distance) between $\mathbf{x_i}$ and $\mathbf{w}_{m,n}$ is the minimum possible among all weights. Subsequently, an iterative process updates the weights at a given learning rate ($h_0$) that decrease over time, and following a neighborhood function that ensures the preservation of the topology. This neighborhood function (usually a Gaussian) is governed by a parameter $\nu$ that defines the initial spread of the neighborhood of each neuron.

The learning process ends after a maximum number of iterations, $n_{max}$, or when the weights do not change significantly \citep{kohonen1982}. Finally, each neuron (and thus the candidates that fell into it) receives the label corresponding to the majority class, taking as a reference the sources with a confirmed classification.

The SOM implementation used in this work is the Python \texttt{MiniSom}\footnote{\url{https://github.com/JustGlowing/minisom/}} library for its ease of use and flexibility in hyperparameter configuration \citep{vettigli2018}. In the following, we will assume a squared map, $M = N$ (for simplicity and because the total number of neurons is much more crucial than their distribution) 

Subsequently, to choose the best hyperparameters for the SOM (namely, the map size $N^2$, $\nu$, $h_0$, and $n_{max}$) we implemented a grid search process, assuming for simplicity a squared map ($N=M$) with $N \in \{5,6,7,8\}$; $\nu\in \left[0.5, 1.5\right]$ and $h_{0}\in \left[0.1, 1.0\right]$, both in steps of $0.1$; and the number of iterations $n_{max}\in \{100, 500, 1000, 5000, 10000\}$.

We built the cost function, $f = f(N, \nu, h_0, n_{max})$ to minimize as a composition of three different metrics: the quantization error (QE), the topographic error (TE), and the $F_1$-score. 

The QE is defined as the mean distance between each element $x_i$ and their BMU and indicates how well the SOM represents the input data \citep{kohonen1982}, while the TE, quantifies the fraction of input samples for which the first and second BMU neurons were not placed adjacents in the map. That is, the TE is a measure of how well the SOM preserved the topology \citep{kiviluoto1996}. Both QE and TE are computed with equations \eqref{QE} and \eqref{TE}:
\begin{equation}\label{QE}
    QE = \dfrac{1}{n}\sum_{i=0}^{n}||\text{BMU}(x_i) - x_i||^2
\end{equation}
\begin{equation}\label{TE}
    TE = \dfrac{1}{n}\sum_{i=0}^{n}\epsilon(x_i),
\end{equation}

where $\epsilon(x_i) = 1$ if the first $\text{BMU}(x_i)$ and the second $\text{BMU}(x_i)$ are not adjacents, and $\epsilon(x_i) = 0$ otherwise.

On the other hand, the $F_1$-score is defined as the harmonic mean of the precision\footnote{The precision of a class is computed as ${TP}/{(TP+FP)}$, being $TP$ the number of True Positives and $FP$ the number of false positives. In simple terms, it provides the probability of the algorithm to be correct when assigning a class to a source,} and the recall\footnote{The recall is computed as ${TP}/{(TP+FN)}$, with $FN$ being the number of False Negatives. It indicates the ratio of sources of a given true class that are correctly classified as that class.} of the classification, and calculated with the equation \eqref{F_1}:
\begin{equation}\label{F_1}
F_1 = 2\times\dfrac{\text{Precision}\times \text{Recall}}{\text{Precision}+\text{Recall}}.
\end{equation}
Therefore, to achieve a good SOM classification, we must aim to minimize $QE$ and $TE$, while maximizing the $F_1$-score. To do this, we expressed $f = max\{\tilde{QE},\, \tilde{TE},\, 1-\tilde{F_1}\}$, where the symbol `` $\tilde{}$ '' means that those three quantities have been previously scaled with the min-max normalization, and look for the minimum in the parameter space shown above.
As a result, we found the global minimum with a map size of $8 \times 8$ neurons, $\nu$ = 1.4, $h_0 = 0.4$, and $5000$ maximum iterations.

\begin{figure}
    \centering
    \includegraphics[width=0.5\textwidth]{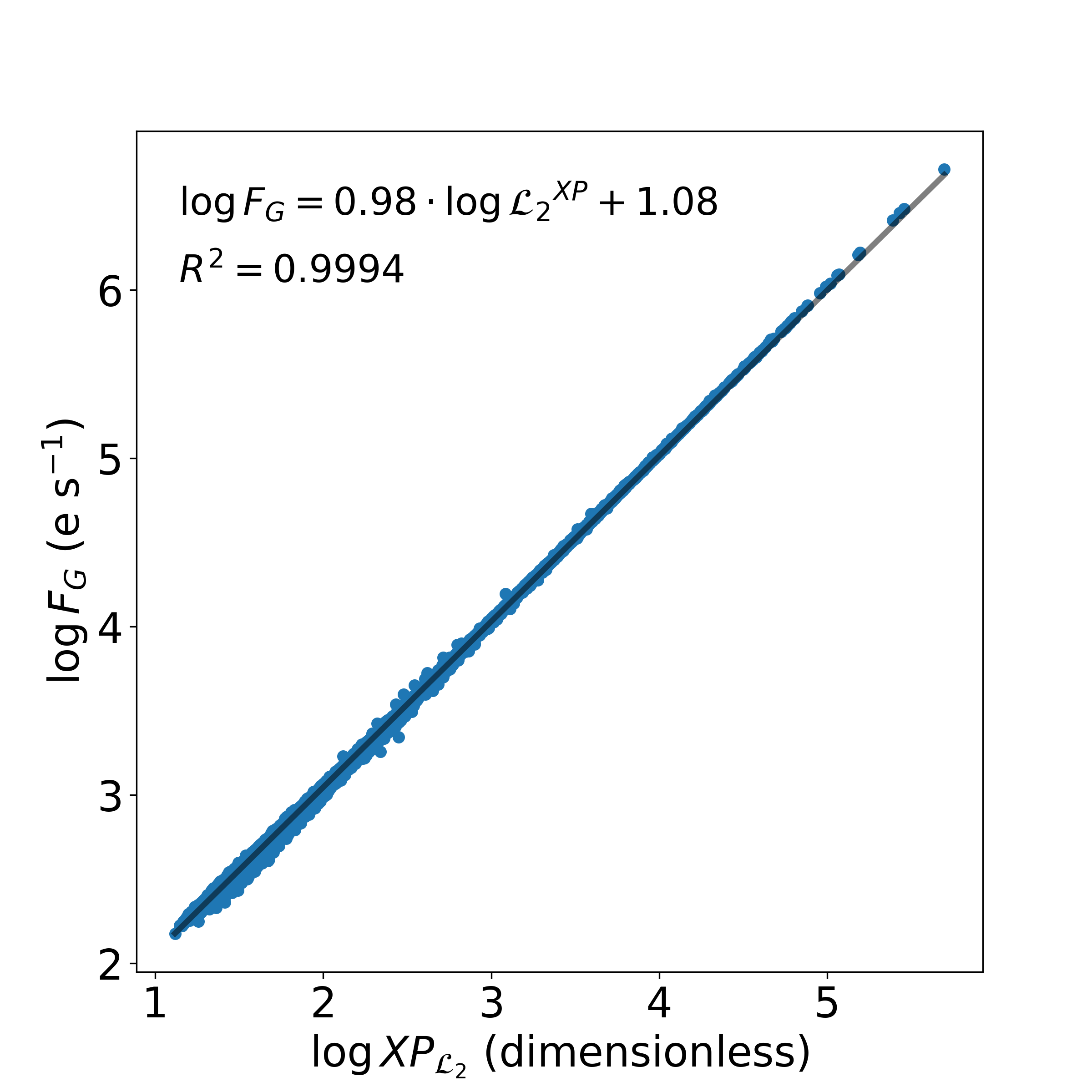}
    \caption{Linear relation between $\mathcal{L}_2^{XP}$ and $F_G$ as calculated with the $90,667$ sources used in this work.}
    \label{fig:XP_FG_corr}
\end{figure}

It is important to recognize that the effectiveness of this approach, like other distance-based algorithms, is strongly tied to the scale of the features involved (here, the XP coefficients). To remove distance-dependent effects and focus the SOM training on spectral morphology rather than flux amplitude, we normalized each XP coefficient vector by its $\mathcal{L}_2$ norm ($\mathcal{L}_2^{XP}$), which linearly correlates with the $G$-mean flux ($F_G$) of the source, as shown in Figure \ref{fig:XP_FG_corr}. This procedure has already been used in \citet{deangelietal2023} to optimize the set of basis functions used to represent the XP spectra.

\section{Results}

\subsection{Spectral classification}

\begin{figure}
    \centering
    \includegraphics[width=0.49
    \textwidth]{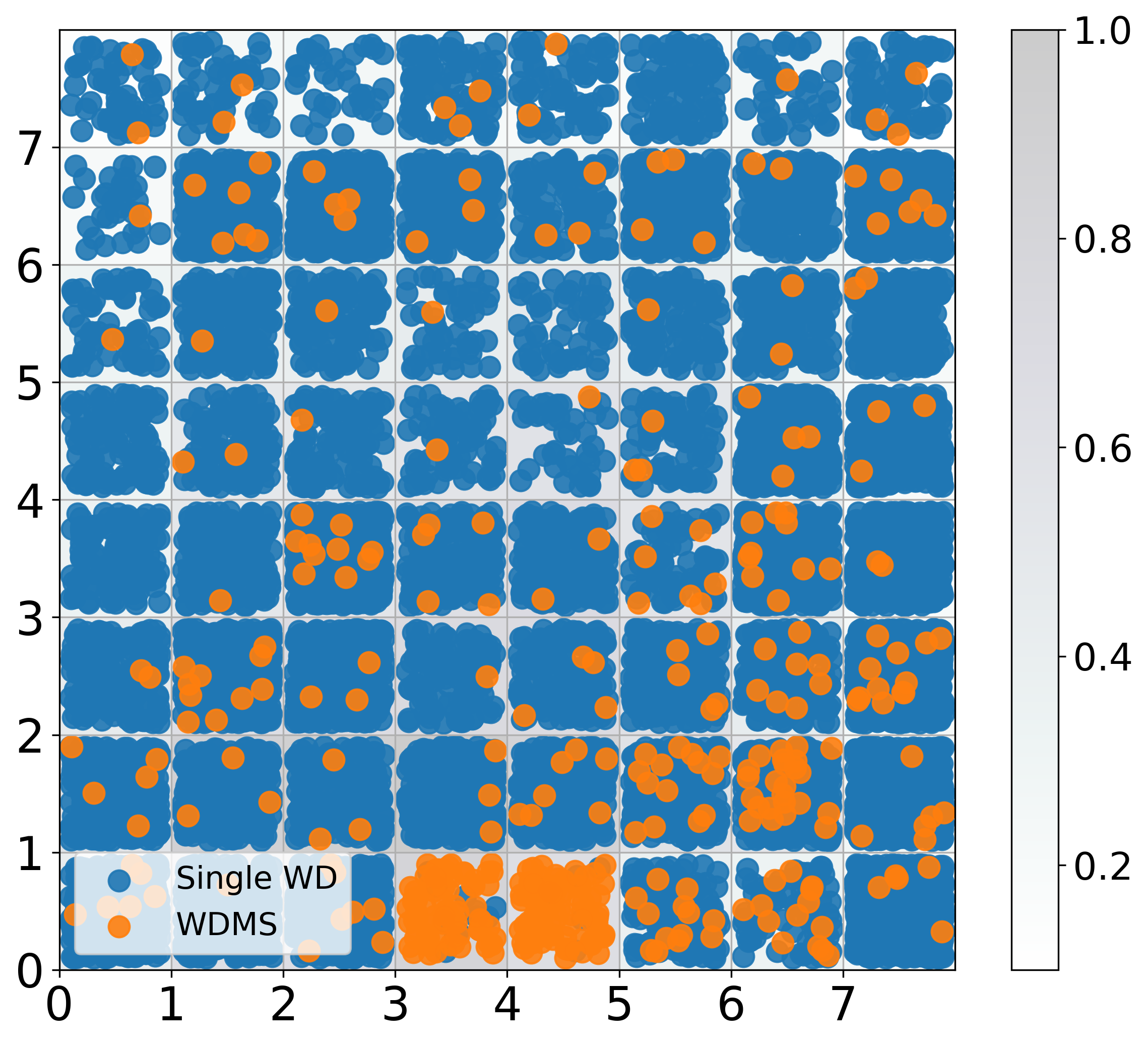}
    \caption{SOM map with our sample of $90,667$ sources. WDs with a confirmed MS companion appear in orange, single WDs in blue, and candidates are invisible to enhance visualization.}
    \label{fig:som_binarity}
\end{figure}

We incorporated the $90,667$ sources in the form of normalized XP data into the SOM with the hyperparameters defined in \S \ref{som_description}. The resulting map is shown in Figure \ref{fig:som_binarity}, where confirmed WDMS binaries are plotted in orange, single WDs in blue, and candidates are invisible to enhance visualization.

As illustrated, some confirmed WDMS binaries share the same neurons as single WDs due to their XP composite spectra being entirely dominated by the WD component. Notwithstanding that, two neurons ($z_{3,0}$ and $z_{4,0}$) are clearly dominated by WDMS binaries. Indeed, among the WDs fallen in neuron $z_{3,0}$, $85\%$ are confirmed WDMS binaries; a percentage that is increased up to $92\%$ in neuron $z_{4,0}$. Therefore, we labeled them as WDMS neurons. The other 23 neurons (having a percentage of WDMS $<50 \%$) are considered in the following as single WD neurons.

\begin{figure}
    \centering
    \includegraphics[width=0.45\textwidth]{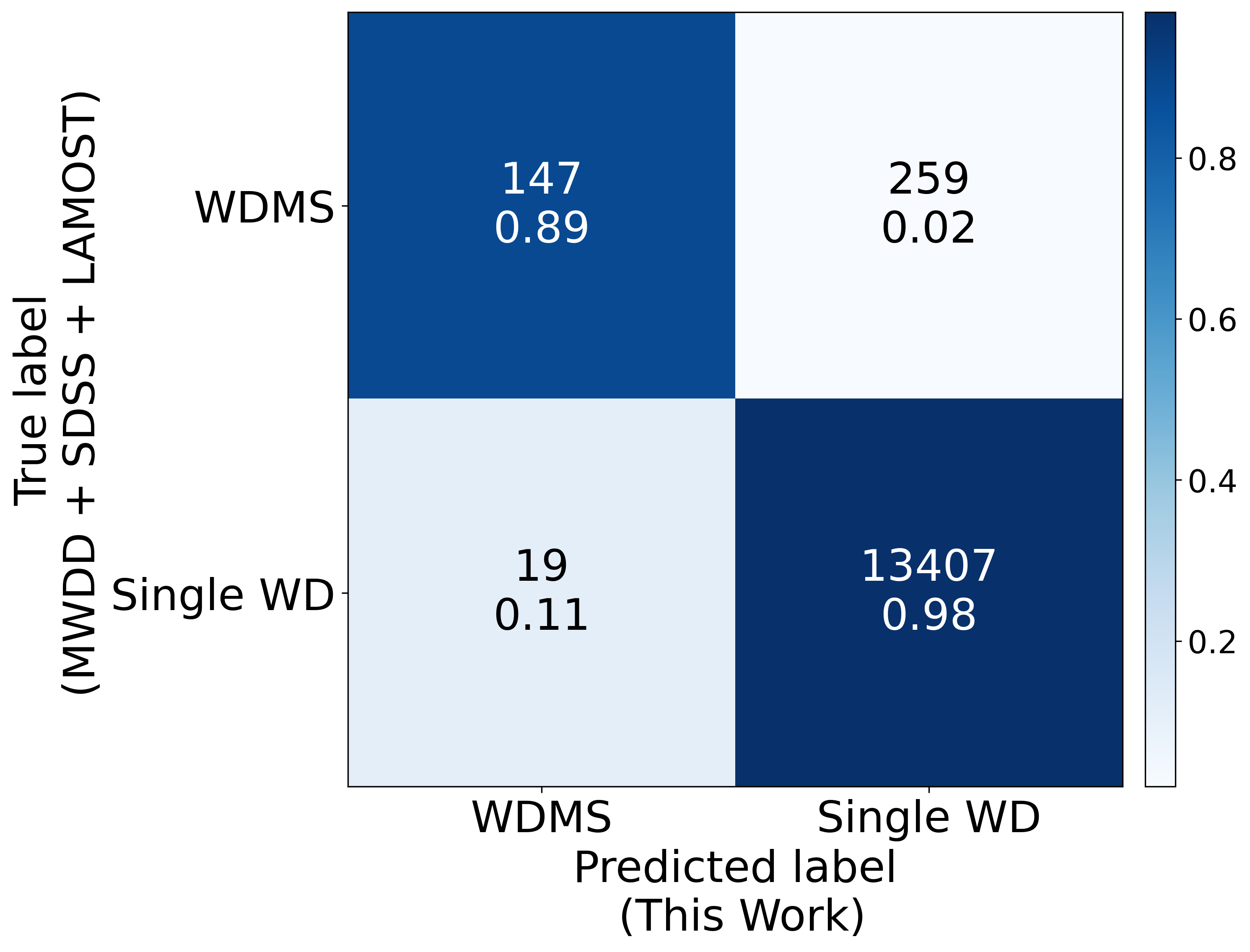}
    \caption{Confusion matrix of the binary WDMS - Single WD SOM classification.}
    \label{fig:WDMS_confusion_matrix}
\end{figure}

Using this labeling procedure, we can compute a confusion matrix (see Figure \ref{fig:WDMS_confusion_matrix}), as well as precision and recall metrics (see Table \ref{tab:som_metrics}) to validate our methodology. The confusion matrix, $C$, has as rows the true labels (that is, those used as a reference, here MWDD combined with SDSS and LAMOST) and as columns the predicted labels (those assigned after the SOM clustering plus the labeling procedure described above). In this way, each cell $C_{i,j}$ contains the number of sources of the $i$ class, classified by our SOM as belonging to the $j$ class. 

In Figure \ref{fig:WDMS_confusion_matrix} we show the confusion matrix with the numbers described above in each cell, and below them the same number normalized by columns, which is equivalent to the precision. In addition to that, in Table \ref{tab:som_metrics} the precision, recall, and $F_1$-score for each class are summarized. 

\begin{table}[ht]
\caption{\label{tab:som_metrics} {Precision and recall metrics for WDMS-single WD classification.}}
\begin{center}
\begin{footnotesize}
\begin{tabular}{lccc}

{Class}  & {Precision} & {Recall} & {$F_1$-score}  \\ \hline
WDMS & 0.89      & 0.36  & 0.51  \\
Single WD & 0.98      & 1.00  & 0.99  \\ \hline
\end{tabular}
\end{footnotesize}
\end{center}
\end{table}
As can be seen, our classification shows excellent precision ($\sim 90\%$) in identifying WDMS binaries. However, its low recall (only a third of the WDMS binaries are classified as such) suggests that they are systems where the WD flux overwhelms that of the cool companion.

Although both $z_{3,0}$ and $z_{4,0}$ contain WDMS binaries, they are different neurons which, based on the conservation of the topology order, suggests that there is some difference between their populations. Indeed, the median, 25th and 75th percentiles of the $G_{BP} - G_{RP}$ color of the input samples in the $z_{3,0}$ neuron (hereafter, the cool WDMS neuron) is $0.47^{+0.09}_{-0.09}$ mag, while for the $z_{4,0}$ neuron (hereafter, the hot WDMS neuron) is $0.12^{+0.08}_{-0.06}$ mag. Moreover, by using the $T_{eff}$ computed in \citet{gentilefusilloetal2021} from the $G$, $G_{BP}$, and $G_{RP}$ photometry assuming $H$-rich atmospheres, we obtained a corresponding median $T_{eff}$ of $\sim 7500^{+800}_{-400}$ K for the cool WDMS neuron and $\sim 11,000^{+1600}_{-1000}$ K.

\begin{figure}[ht]
    \centering
    \includegraphics[width=0.5
    \textwidth]{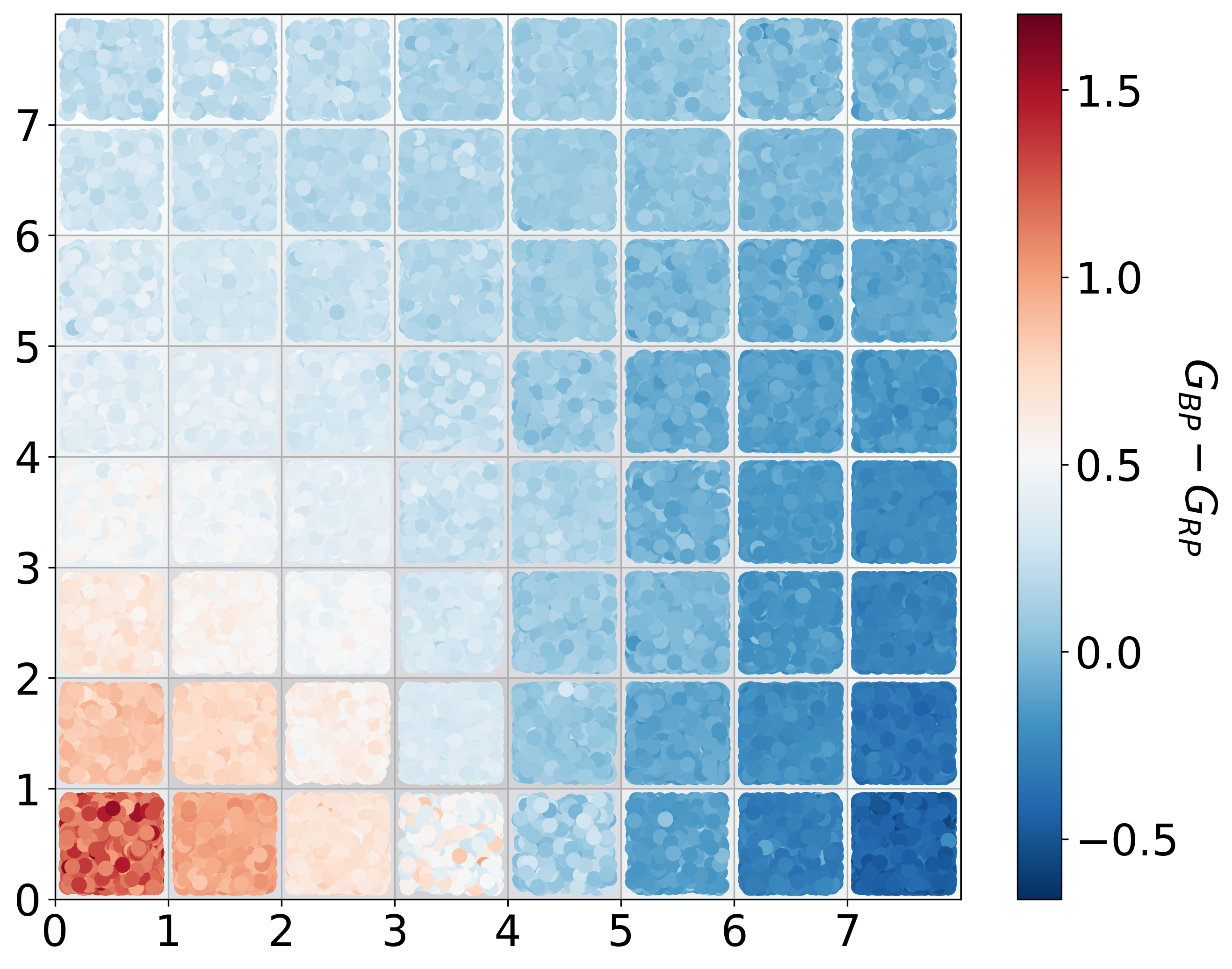}
    \caption{The SOM map displays the $G_{BP}-G_{RP}$ color of the $90,667$ sources. A smooth, non-linear $G_{BP}-G_{RP}$ color gradient is shown.}
    \label{fig:som_temperature}
\end{figure}

In general, the distribution of $G_{BP}-G_{RP}$ color across the two axes of the SOM is not expected to be irregular, since color and, correlatively, the $T_{eff}$ are highly dependent on the spectral shape. Indeed, if we plot the $G_{BP}-G_{RP}$ color of the $90,667$ sources present in the SOM, we see a smooth, non-linear gradient with the bluer sources on the left and the cooler ones on the right, as shown in Figure \ref{fig:som_temperature}.

There are 993 sources classified as WDMS binaries (525 in the cool WDMS neuron and 468 in the hotter one), of which 846 ($85 \%$) have not yet been classified as WDMS binaries in the MWDD, SDSS, or LAMOST catalogs. These sources are therefore new WDMS binary candidates. 

\begin{figure*}[ht]
\gridline{\fig{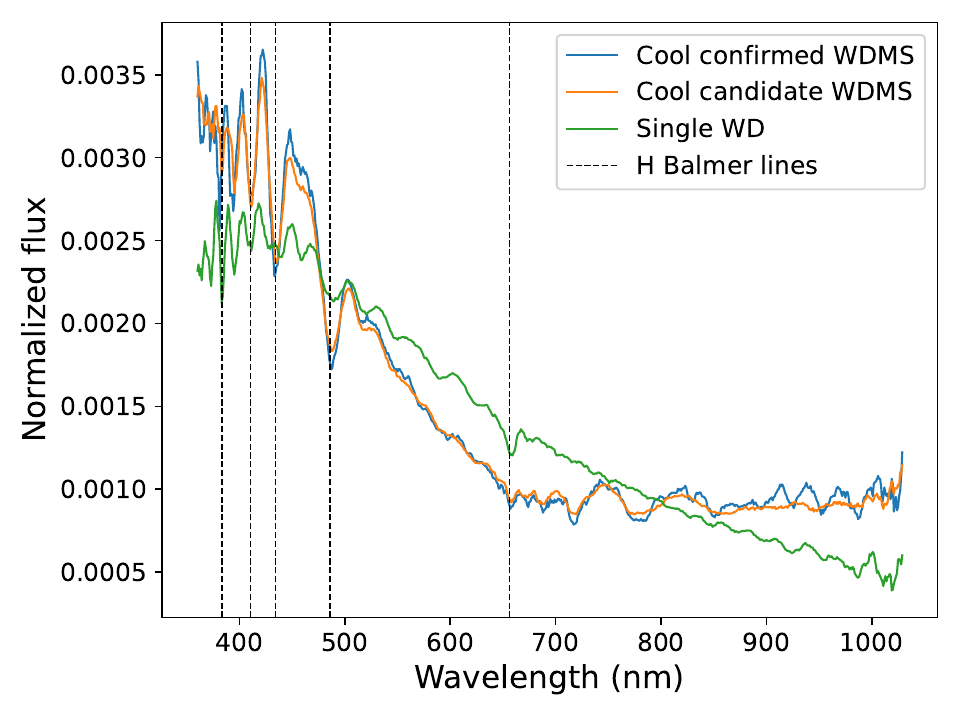}{0.495\textwidth}{(a) Cool WDMS candidates median spectra (orange) vs cool confirmed WDMS median spectra (blue).} \label{cool_wdms_median_spectra}
\fig{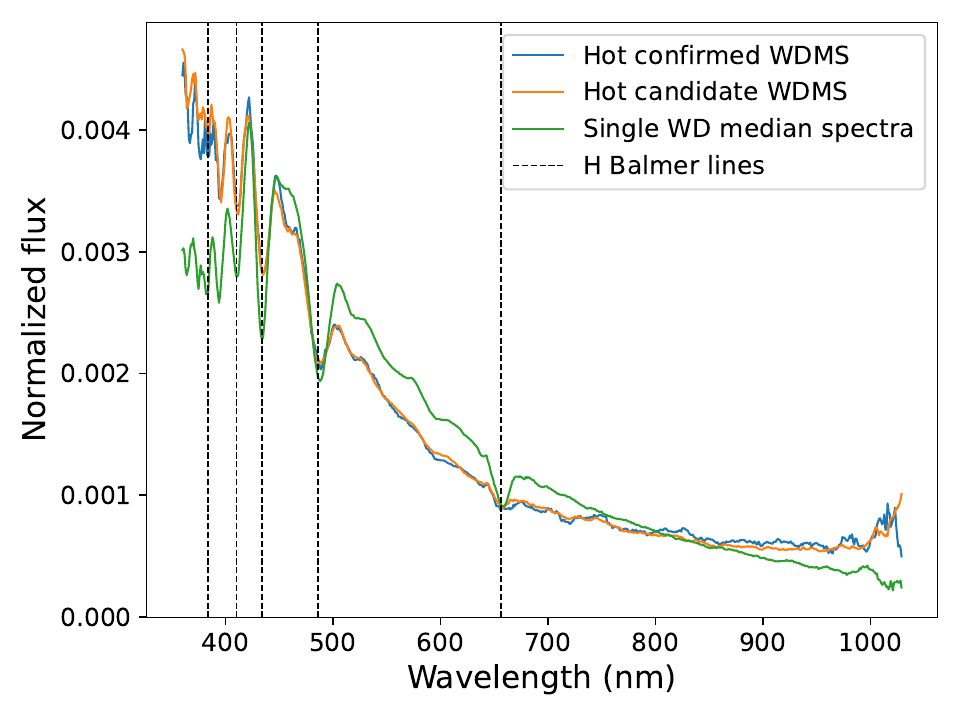}{0.495\textwidth}{(b) Hot WDMS candidates median spectra (orange) vs hot confirmed WDMS median spectra (blue).} \label{hot_wdms_median_spectra}
}
\caption{\label{median_spectra} Comparison between the normalized median spectra of the confirmed WDMS (blue) and that of the candidates (orange) for both the cool and hot WDMS neurons. The median spectra of a single WD neuron with comparable $G_{BP}-G_{RP}$ color is included (green) so that the red flux excess can be seen.}
\end{figure*}

In Figures \ref{median_spectra}a and \ref{median_spectra}b we show the normalized median externally-calibrated spectra of the cool and hot WDMS neurons, obtained with the \texttt{GaiaXPy}\footnote{\url{https://gaia-dpci.github.io/GaiaXPy-website/}} library. For comparison, we also show in green the normalized median spectra of single WD neurons, with a median $G_{BP}- G_{RP}$ color similar to that of the WDMS neurons, so that the continuum can be compared.

As can be seen in Figures \ref{median_spectra}a and \ref{median_spectra}b, both cool and hot WDMS median spectra show a clear red flux excess with respect to the single WD continuum, thus indicating the presence of an optical, late-type stellar companion in their composite spectra. 

There exists the possibility that a red flux excess is due to the emission from a disk around the WD \citep{melisetal2012, brinkworthetal2012, farihietal2012, xuandjura2012, hartmannetal2016, rogersetal2024, swanetal2024}. We explored the possibility that the map mistook cool companions for hot disks. We put the sample of 33 WDs with disks recorded so far in the MWDD and with available XP spectra, and found that none of them fell into the binary neurons. While this does not fully exclude the possibility that some of our WDMS pairs are WDs with disks, we take as a working hypothesis that those red flux excesses are associated with MS stars due to the low number of disks observed surrounding WDs \citep[about $1-3\%$,][]{wilsonetal2019}.

Furthermore, to assess the reliability of our morphological clustering, we have compared the median spectra of the 455 (391) cool (hot) WDMS binary candidates with that of the 70 (77) cool (hot) confirmed WDMS binaries in each neuron. As shown in the same Figure \ref{median_spectra}, in both cases the median spectra of the confirmed and candidate WDMS binaries overlap almost perfectly.

\subsection{WDMS eclipsing binary candidates}\label{sec:eclipsing}

It is of great interest to look for variability indicators in our 993 sources' sample, since it seems reasonable to expect that the orbit of some of those systems could be aligned with the line-of-sight of Gaia, turning them into eclipsing binaries.

% As commented on Section \ref{other_methods} our WDMS candidates mainly fall in the \citet{kaoetal2024} UMAP region with the highest probability of containing eclipsing binaries. 
Indeed, 101 ($10\%$) of our 993 WDMS candidates appear as variable sources in the Gaia Archive ($\texttt{phot\_variable\_flag = "VARIABLE"}$) so it is tempting to link that variability with the binarity clues found in their XP spectra, either because they may be CVs or eclipsing binaries. This fact is particularly enlightening given that merely 1651 sources ($2\%$) are found to be variable within those sources in the single WD neurons.

To shed more light on this issue, we cross-matched our sample with the all-sky Gaia DR3 Eclipsing Binary catalog \citep[\texttt{gaiadr3.vari\_eclipsing\_binary} table, see][]{mowlavietal2023} that contains $2,184,477$ eclipsing binary candidates obtained from $G$-band light curves cleaned and modeled to find their orbital period.

As a result, we found that 13 ($1\%$) of our WDMS binary candidates appear in that catalog. In contrast, 100 times fewer eclipsing binary candidates are found in the single WD neurons: only 13 sources, or $0.01\%$. The orbital periods ($P$) of our 13 WDMS eclipsing binary sample are available, ranging from $\sim 0.2$ to $\sim 1.5$ days, with a median of 0.5 days. This finding suggests that their orbits are particularly close.

\subsection{Stellar parametrization with VOSA}\label{vosa_section}

In order to validate our sample of WDMS candidates with external data and to estimate their astrophysical parameters, we used VOSA. This VO tool enables us to gather photometric data from major multi-wavelength astronomical surveys and to compile these data into an observational SED \citep{bayoetal2008}.

This SED is subsequently used to fit the stellar parameters of the source by using any theoretical model publicly available in the literature. Furthermore, VOSA enables the implementation of a binary fit algorithm, which aims to fit two models to the SED simultaneously: one model for each companion. 

To apply VOSA to our WDMS candidates, we used the following input parameters: the Gaia DR3 source ID, equatorial coordinates in J2000.0 (to calculate them from the Gaia J2016.0 epoch, we employed proper motions and parallax information in Gaia, and the \texttt{astropy} library \citep{astropy}), geometric distances from the catalog of \citet{bailerjonesetal2021} and the mean visual extinctions $A_v$ calculated in \citet{gentilefusilloetal2021}.

As source catalogs for the photometric points we used the GALEX GR6/7 \citep{bianchietal2017} in the UV range; SDSS DR12 \citep{alametal2015}, Gaia DR3 \citep{gaiacollaboration2023}, and Pan-STARRS DR2 \citep{magnieretal2020} in the optical; and DENIS \citep{epchteinetal1994}, 2MASS \citep{skrutskieetal2006}, and CatWISE2020 \citep{maroccoetal2021} for the near IR (NIR). Furthermore, since we have Gaia XP spectra available for every source, we also incorporated their J-PAS synthetic photometry with \texttt{GaiaXPy} \citep{benitezetal2014, montegriffoetal2023}. All photometry was programmatically retrieved with VOSA, except that from CatWISE2020 and J-PAS since they are not currently included in VOSA, so we loaded them manually.

Once VOSA has obtained the photometric points of each source, it automatically rejects those points bearing bad quality flags in their respective catalogs. An equivalent procedure was applied to our CatWISE2020 photometry by imposing high-quality flags  ($ccf = 0000$ and $ab\_flag = 00$, see \citet{maroccoetal2021} for further details). Moreover, we discarded any point in the overall photometry with a relative error for the flux greater than $20\%$, and retained only those sources with at least a point from 2MASS and CatWISE2020 photometry, to ensure NIR coverage. This last filter is highly conservative and reduces our final sample of WDMS with computed parameters (from 993 to 323 sources). However, we consider it crucial if we want to obtain a reliable SED fit since the low-mass MS companions are expected to have their emission peak in the NIR. Furthermore, by doing so we prevent overfitting issues due to the high number of optical points mainly provided by the J-PAS synthetic photometry.

Subsequently, we fitted the resulting photometry to three distinct types of models: a single-body fit to the BT-Settl-CIFIST model \citep{baraffeetal2015} (setting $1200 \leq T_{eff}/K\leq 7000$ and $4\leq \log{g}/dex\leq 5$ for MS stars); a single-body fit to the WD Koester model \citep{koester2010} ($5000 \leq T_{eff}/K\leq 80,000$, and $6.5\leq \log{g}/dex\leq 9.5$); and a two-body fit using both models simultaneously. 

It should be noted that, although the WD Koester model assumes hydrogen-rich (DA) atmospheres, this hypothesis is more than reasonable in our work since our WDMS candidates clearly show Balmer lines, as can be seen in Figure \ref{median_spectra}.

To assess the quality of a fit, VOSA uses the visual goodness- of-fit (Vgf$_b$), a modified version of the reduced $\chi^2$ in which the relative photometric errors are considered to be at least $10 \%$, to prevent any underestimation of the uncertainties. In this way, a SED is considered well-fitted if Vgf$_b < 10-15$. Notwithstanding that, \citet{nayaketal2024} have detected that some SEDs fittings with low Vgf$_b$ are not always satisfactory. Moreover, a preliminary analysis of some fits in this work has shown that a Vgf$_b<10$ is compatible with a $\chi^2_{red}$ as high as 100 or 1000. Consequently, we decided to use $\max \{ \chi^2_{red}, \text{Vgf}_b \}<10$ as a more conservative but reliable criterion to define a good quality SED fitting.

From the sample of 323 WDMS binary candidates with available optical and NIR photometry, 137 of them shown an excellent fit to the binary WD Koester -- BT-Settl SED model, according to the criteria described above. Moreover, none of our sources shown a good fit to the single BT-Settl SED, and only one source fitted well to the single WD Koester SED, with a better $\chi^2_{red}$ and Vgf$_b$ than for the binary SED fit, so we discarded it. 

As a result, we have obtained a golden sample of 136 high-confidence WDMS binary candidates for which VOSA provides the best-fitted $T_{eff}$ and bolometric flux ($F_{bol}$) for each companion.

\begin{figure*}[ht]
\gridline{
\fig{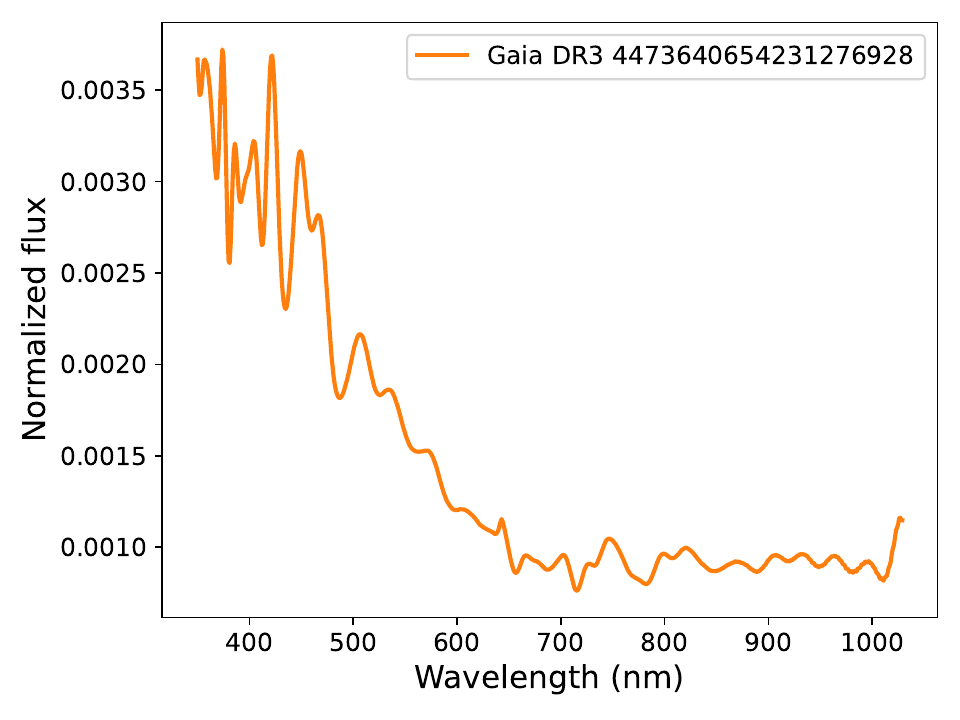}{0.45\textwidth}{(a) Externally calibrated XP spectra of the cool WDMS prototype}
\fig{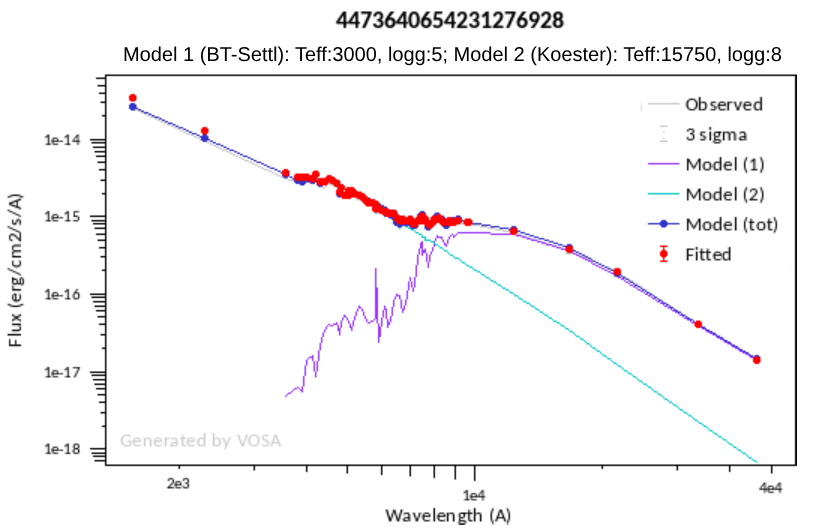}{0.54
\textwidth}{(b) Photometric points found by VOSA for the cool WDMS prototype and fitted to a binary SED}
}

\gridline{
\fig{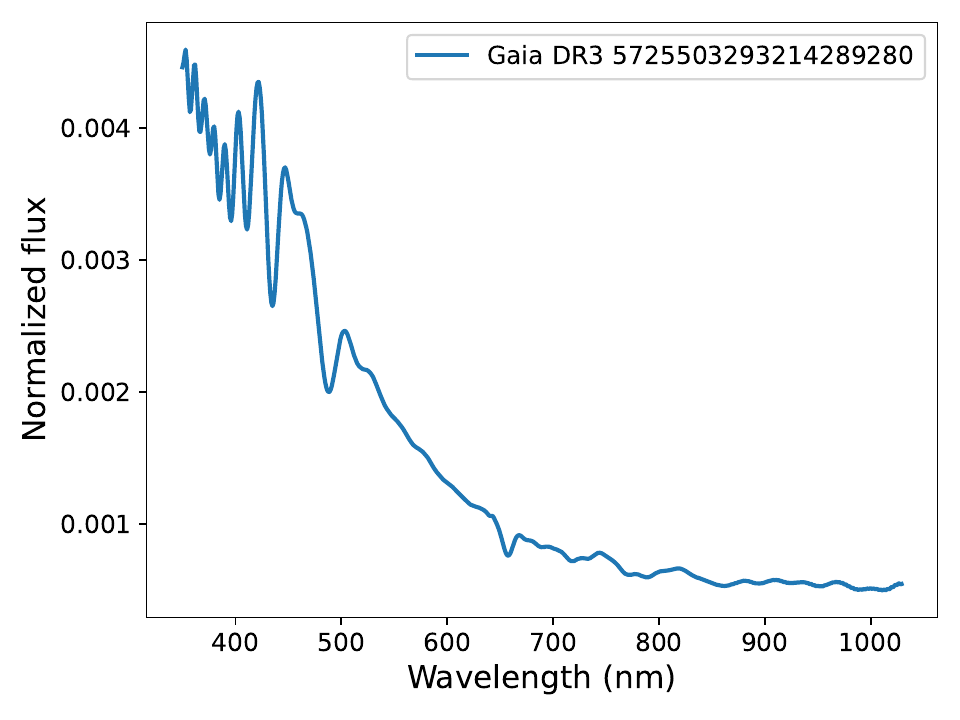}{0.45\textwidth}{(a) Externally calibrated XP spectra of the hot WDMS prototype}
\fig{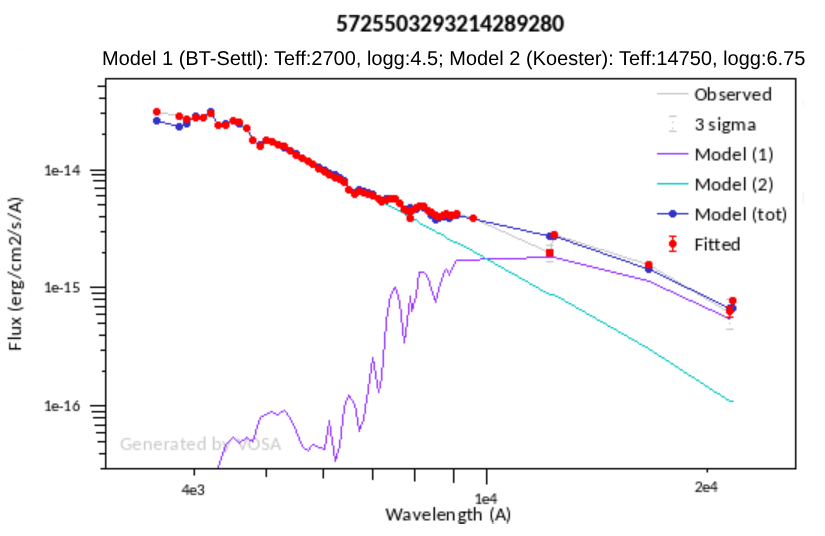}{0.54
\textwidth}{(b) Photometric points found by VOSA for the hot WDMS prototype and fitted to a binary SED}
}

\caption{\label{cool_WDMS_prototype} Calibrated Gaia XP spectra and two-body SED fitted for the cool (top) and hot (bottom) prototypes of the WDMS binary candidates.}
\end{figure*}

In Figure \ref{cool_WDMS_prototype} we present the calibrated XP spectra of the cool and hot WDMS neuron prototype (i.e., the source most similar to the externally-calibrated median spectra) in the left, and their VOSA binary fitted SEDs in the right.

The final WDMS binary candidates show a median, 25th, and 75th $T_{eff}$ percentiles for the WD companion of $15,000^{+3750}_{-2500}$ K, although there is a slight difference between the median $T_{eff}$ of hot and cool neurons ($12,500$ K for the cool neuron and $17,250$ K for the hotter one). These values are approximately $5000-6000$ K higher than those obtained from the $T_{eff}$ calculated in \citet{gentilefusilloetal2021} assuming a single WD. This discrepancy is most likely due to the fact that they only used the $G$, $G_{BP}$, and $G_{RP}$ colors to fit their atmospheric models while we used a significantly larger set of photometric points spanning a wider wavelength range from the UV (where the emission peak in WDs is located) to the NIR.

Regarding the MS companion, the $T_{eff}$ has a median, 25th, and 75th percentile of $2800^{+200}_{-100}$ K that, when translated to spectral types using the updated tables of \citet{pecautandmamajek2013}, is equivalent to a median  M6V type. 

It is worth mentioning that there are 9 sources in which the faint companion has $T_{eff} \leq 2250$ K, compatible with a brown dwarf (BD) candidate \citep{pecautandmamajek2013, kirkpatricketal2021}. Further spectroscopic follow-up observations are planned to confirm these objects.

Finally, we leveraged VOSA to fit the 406 confirmed WDMS binaries that were used during the labeling process of the SOM training to a binary SED, using the same models as above (but without the NIR photometry requirement). We found, with $max\{\chi^2_{red},\,\text{Vgf}_b\}<10$, that the median $L_{bol}$ ratio (computed as $L_{bol,MS}/L_{bol,WD}$) for the missed WDMS binaries is 0.02, while that of the detected WDMS is 0.13, and their difference is statistically significant (Mann-Whitney U's test p-value $\approx 10^{-14} \lll 0.05$). This confirms that the WDMS binaries missed by the SOM are those whose WD flux overwhelms that of the cool companion.

\subsection{Stellar masses}\label{subsec:stellar_masses}

In principle, WD masses can not be directly determined by VOSA, since the SED fitting has not enough sensitivity to $\log{g}$ which, furthermore, it has an uncertainty as large as 0.5 dex. Therefore, to compute the WD mass ($M_{\text{WD}}$) we used the evolutionary models of \citet{bedardetal2020} along with the $T_{eff}$ and $L_{bol}$ determined in the previous Section \S\ref{vosa_section}.

\begin{figure}
    \centering
    \includegraphics[width=0.58\textwidth]{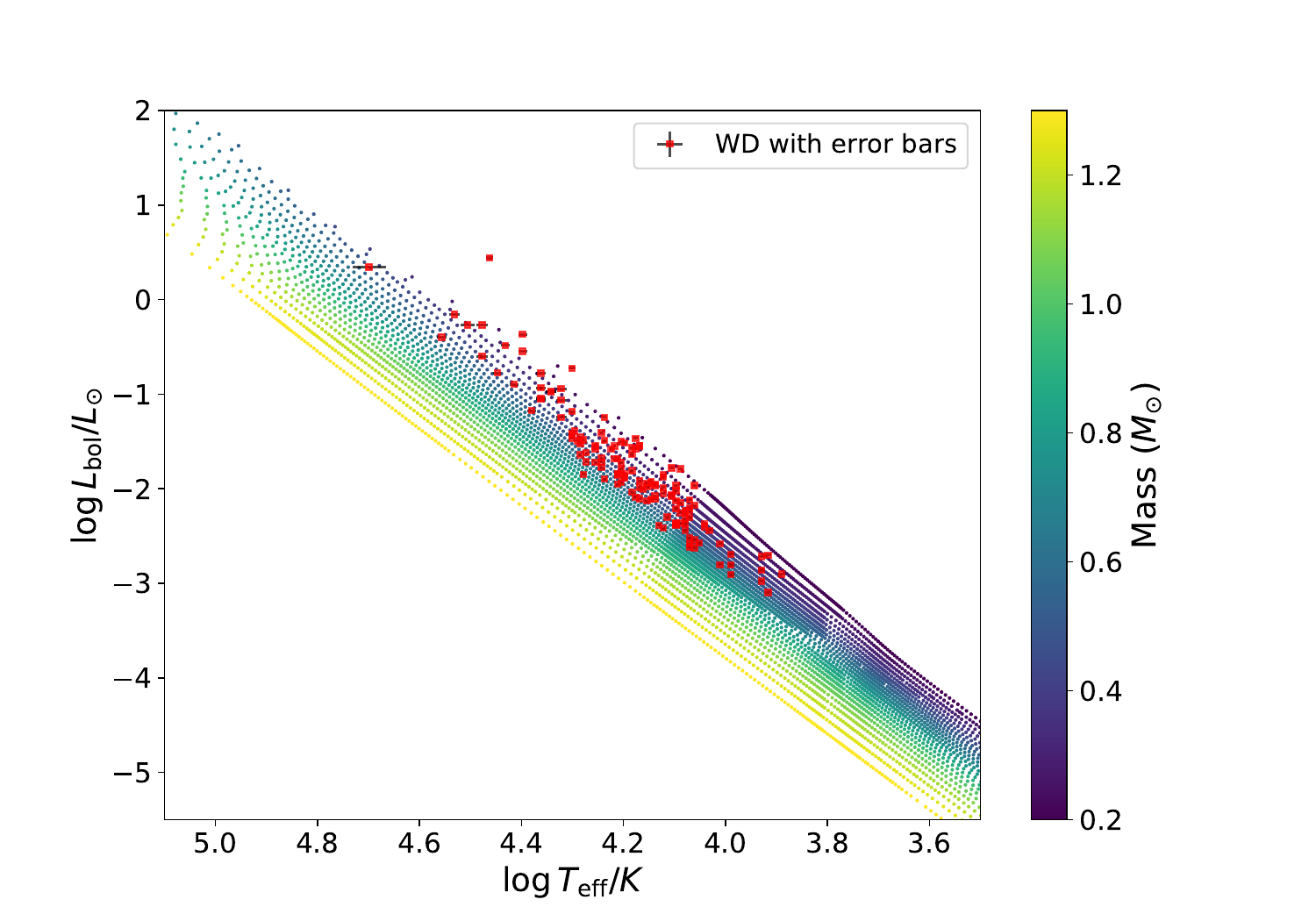}
    \caption{WD evolutionary sequences from \citet{bedardetal2020} and the WD companions of this work}
    \label{fig:evol_seq}
\end{figure}

In Figure \ref{fig:evol_seq} we present the WD evolutionary sequences of \citet{bedardetal2020} in a $L_{bol} - T_{eff}$ diagram for different masses (among $0.2 M_{\odot}$ and $1.3 M_{\odot}$), assuming a C/O core, He mantle, and a thick H outer layer. Over them, we plotted our 136 golden WDMS binary candidates.
Subsequently, we calculated the $M_{\text{WD}}$ for each WD by means of a linear interpolation. There are only three sources out of the convex hull of the WD evolutionary tracks, indicating that they may possess masses lower than $0.2 M_{\odot}$. However, we decided to not compute $M_{\text{WD}}$ for them to avoid extrapolated values. Moreover, we interpolated the $T_{eff}$ and $L_{bol}$ upper and lower values to obtain upper and lower limits of the mass.

As a result, we found that the remaining 133 WD companions have masses ranging from 0.20 to 0.77 $M_{\odot}$, with a median, 25th, and 75th percentile of $0.41^{+0.09}_{-0.08}$ $M_{\odot}$, thus revealing a population of very-low-mass WDs. Indeed, 26 sources (20$\%$ of the sample) have $< 0.3$ $M_{\odot}$ and are therefore considered extremely low mass (ELM) WDs.

Some authors have suggested that the ELM WDs are likely part of Post-Common Envelope Binaries (PCEBs), according to recent studies that found a substantially larger fraction of low-mass WDs in close binaries than in wide binaries. In fact, the latter show a mass distribution similar to that of single WDs (see \citet{mansergasetal2011} and references therein). According to that hypothesis, low-mass WDs are expected to have suffered mass transfer episodes during their Red Giant Branch (RGB) phase in which its companion cannibalized the WD progenitor, penalizing the final mass of the WD.

Regarding to the MS companions, we estimated their masses ($M_{\text{MS}}$) by means of a cubic spline interpolation of their $T_{eff}$ through the \citet{pecautandmamajek2013} tables.

Using the masses $M_{\text{WD}}$ and $M_{\text{MS}}$ of four of the WDMS eclipsing binary candidates for which the orbital period is known (see \S \ref{sec:eclipsing}) we obtained their semi-major axes, $a$, using the Kepler's Third Law: $a = \sqrt[3]{\left(M_{\text{WD}} + M_{\text{MS}}\right){P^2}}$. Not surprisingly, they were found to be very small, with $a$ ranging from $\sim 0.01$ to $\sim 0.03$ au. 

These results points out again toward the PCEB hypothesis, according to which the WD progenitor and its low-mass companion must have been in a relatively tight orbit for mass transfer to occur. If the secondary can't hold the extra material, both stars could be enveloped in a CE through Roche lobe overflow \citep{willemsandkolb2004}.

Within such an envelope, drag forces are expected to remove a significant fraction of the system’s angular momentum, leading to further orbital contraction. The end result would likely be a very close binary system consisting of a low-mass WD and a faint companion star whose mass is too small to produce any noticeable signatures in the system’s astrometry or photometry, which is in keeping with what we observe.

\subsection{Comparison with previous works}\label{other_methods}

To gain some insight into how many of our WDMS candidates are indeed new identifications, we compared them with the 100 pc volume-limited sample of Gaia EDR3 WDMS binaries from \citet{mansergasetal2021b} (hereafter, RM21), finding three common sources. \citet{nayaketal2024} (hereafter, N24) used the Gaia CMD but in combination with UV data from GALEX GR6/7, and identified 93 WDMS, two of which are in our sample, but also in RM21. None of our sources are in the catalog of WDMS binaries in open clusters of \citet{grondinetal2024}.

We found that 157 sources are in common with the work of \citet{lietal2025} where the authors used a supervised ML technique known as Gaussian Process Classifier trained with synthetic data to identify WDMS binaries using the XP spectra of sources among 10 million stars within 1 kpc.

Furthermore, we compared our results with the work of \citet{kaoetal2024}, where a UMAP allowed the authors to project the XP spectra of the high-confidence WDs from the \citet{gentilefusilloetal2021} catalog, in a two-dimensional manifold where similar elements fall close to each other. Subsequently, they used the RUWE and a photometric scatter metric to trace the most likely position of the WDMS binaries in the UMAP. As a result, they found an island of 1096 WDMS candidates.

\begin{figure}
    \centering
    \includegraphics[width=0.47\textwidth]{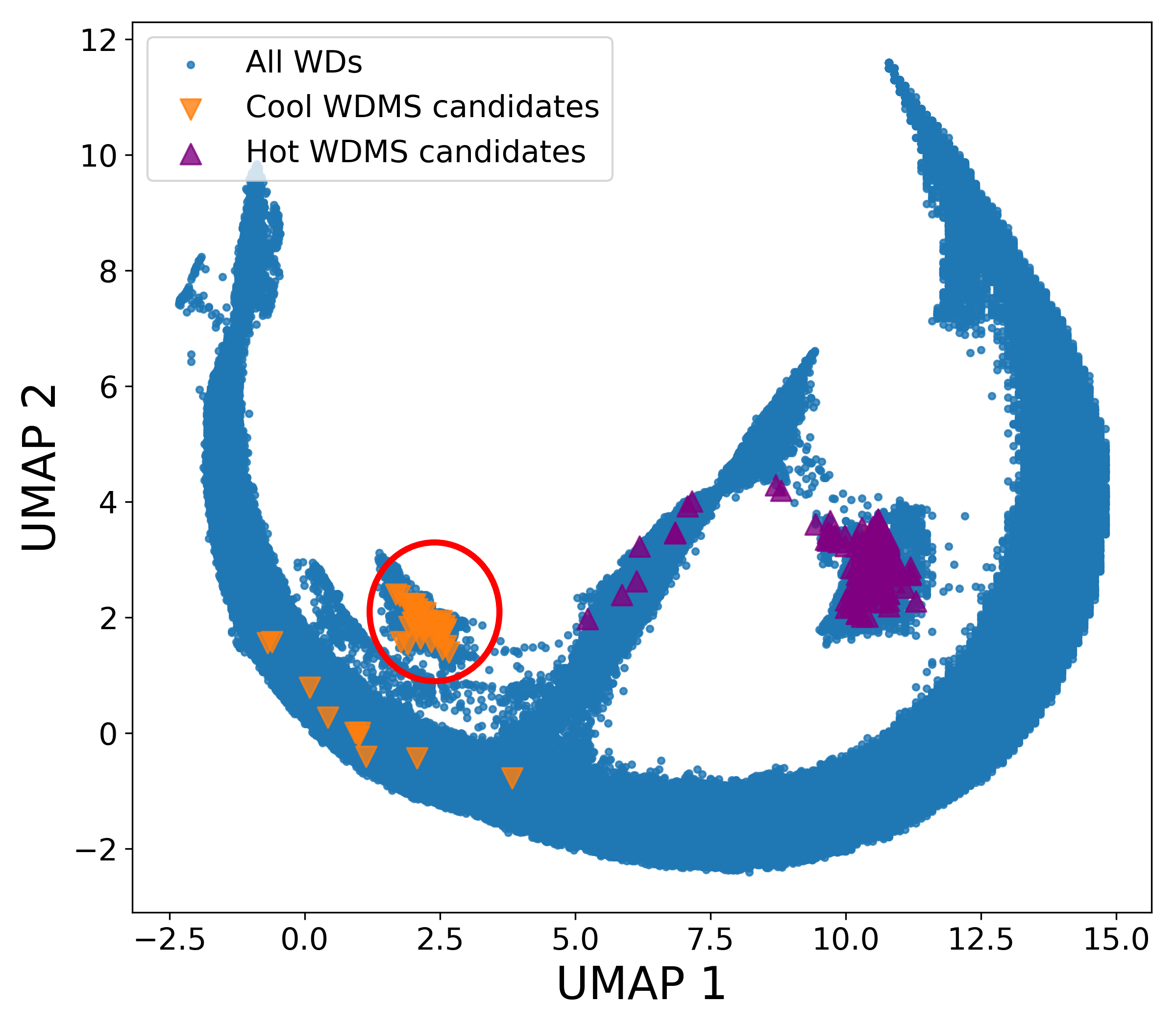}
    \caption{UMAP of \citet{kaoetal2024} with WDs in blue and our WDMS binaries in orange. The WDMS island identified by the authors is delineated as a red circle.}
    \label{fig:umap}
\end{figure}

After finding that 368 of our WDMS are in their catalog, we plotted them in their UMAP as orange and purple triangles (corresponding to sources of our cool and hot WDMS neuron, respectively), as can be seen in Figure \ref{fig:umap}. The WDMS island found by \citet{kaoetal2024} is inside the red circle, where 42 sources of our cool WDMS neuron fell. Not surprisingly, the sources of the hot WDMS neuron are located on the opposite side of the UMAP.

It should be noted that our WDMS candidates are located in the regions of the UMAP with higher values of RUWE and photometric scatter (see Figure 5 in their work), demonstrating a strong agreement between the present study and their work.

During the review period of this paper, \citet{mansergasetal2025} published a new version of their previous work. The authors presented a magnitude-limited catalog of WDMS binaries following a similar methodology that in their previous work \citep{mansergasetal2021b} to filter the sources. However, this time they were not limited to a specific volume and incorporated synthetic photometry from Gaia XP spectra to improve the quality of their VOSA fits. As a result, \citet{mansergasetal2025} (hereafter, RM25) published a larger catalog of 1312 WDMS, of which 356 are in common with our work. This comparison allows us to report that 506 of our sources are totally new identifications.

\begin{figure}
    \centering
    \includegraphics[width=0.4
    \textwidth]{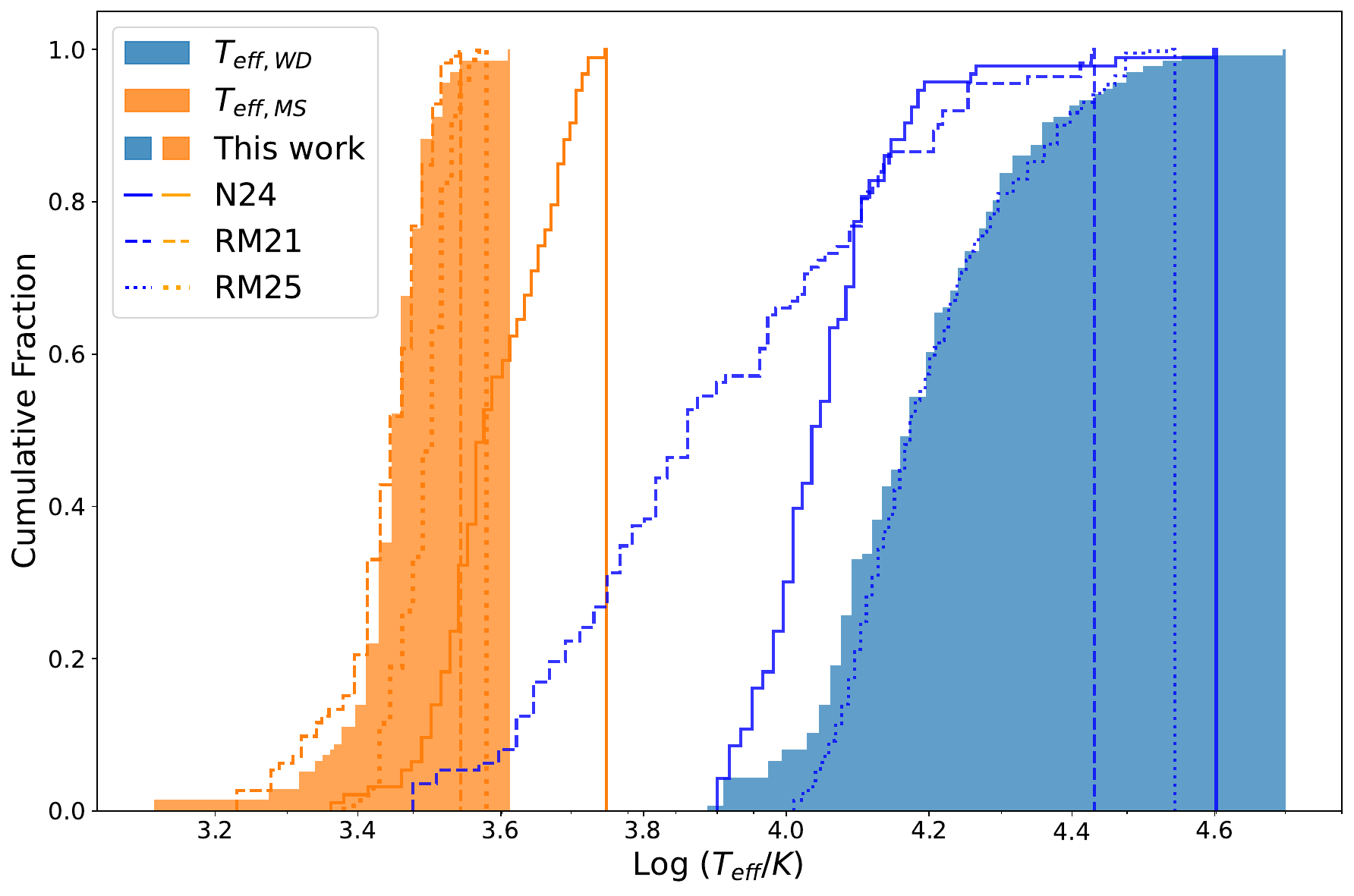}
     \includegraphics[width=0.4
    \textwidth]{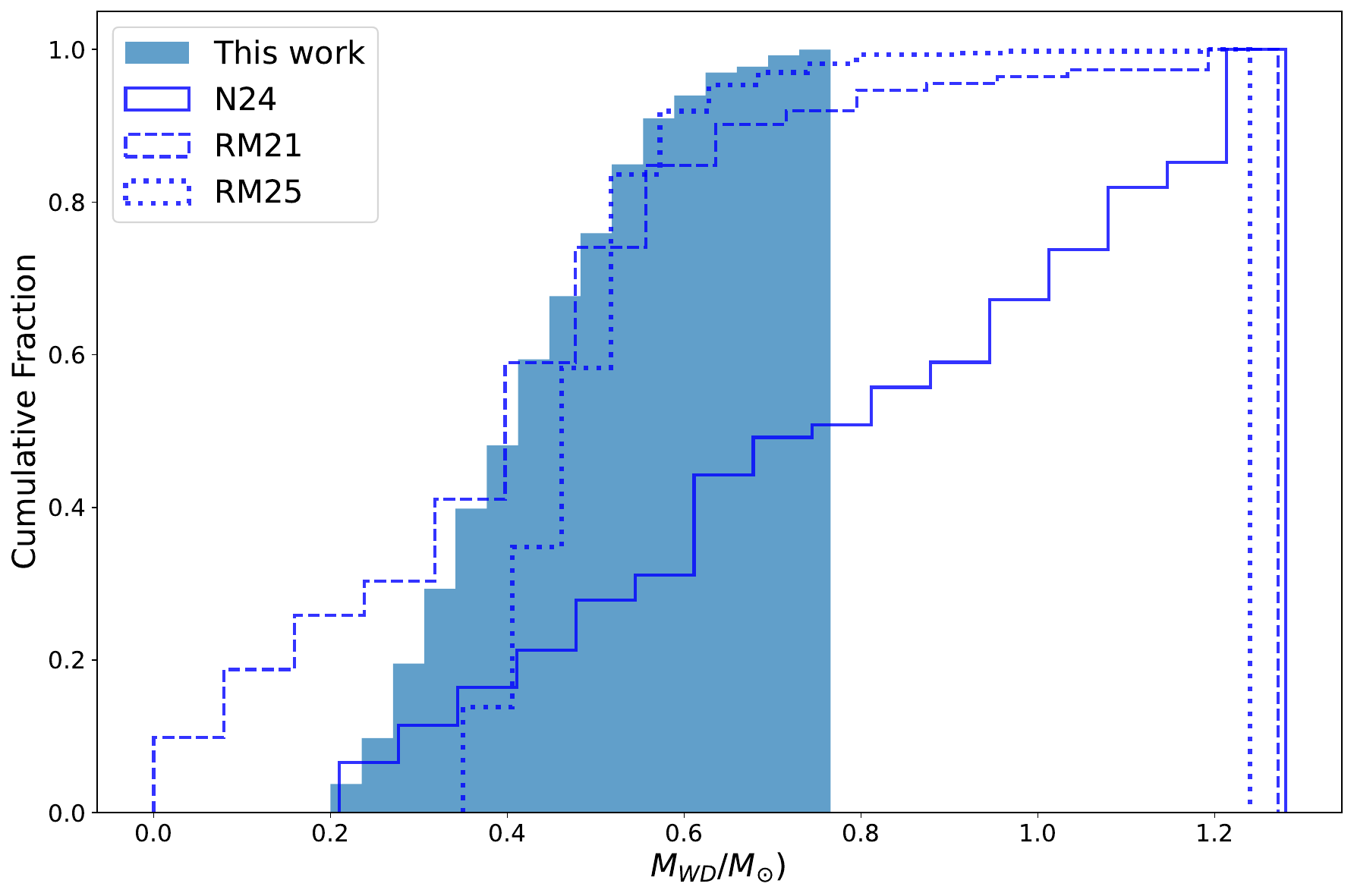}
    \includegraphics[width=0.4
    \textwidth]{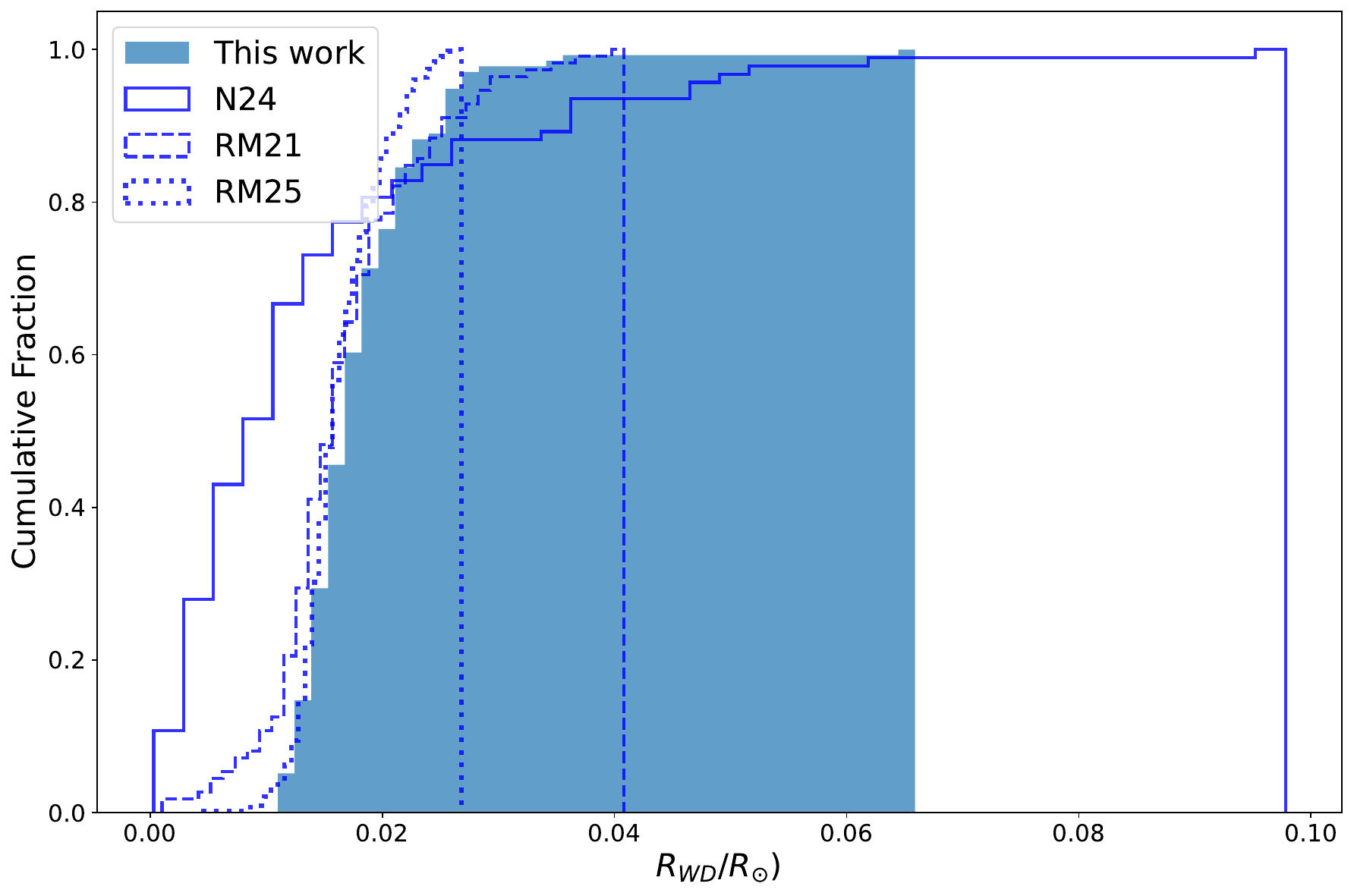}
    \caption{Cumulative $T_{eff}$ distribution for each companion in our golden sample (top), cumulative $M_{\text{WD}}$ distribution (middle), cumulative $R_{WD}$ distribution (bottom) and comparison with the corresponding distributions in RM21 and N24.}
    \label{fig:massradiusteff_distribution}
\end{figure}

Finally, we have compared the WD mass, radius, and $T_{eff}$ of both companions of our golden WDMS candidates with those obtained in RM21, N24, and in \citet{mansergasetal2025} (hereafter, RM25) as shown in Figure \ref{fig:massradiusteff_distribution}.

As can be seen, WDs in our sample are hotter than those in N24 and RM21, although their masses and radii are quite similar, in average, to those in RM21 (though concentrated in a smaller range) thus indicating that our sample belongs to a younger WD population. Moreover, they are lighter and bigger than those found in N24. Furthermore, the cumulative distributions show that our WD primaries are quite similar to those in RM25, with slightly lower masses.

Regarding our MS companions, they are cooler than those in N24 and more similar to those found in RM21 and RM25, except for a small group in the left corner of the distribution, corresponding to the BD candidates found in Section \ref{vosa_section}. This shows that those exotic sources are mostly excluded from previous works.

\setlength{\tabcolsep}{2pt}

\begin{table}[]
\begin{center}
\begin{footnotesize}
\caption{Median, 25th, and 75th percentile of some stellar parameters obtained for our sample in comparison with the works of and \citet{mansergasetal2021b}, \citet{nayaketal2024}, and \citet{mansergasetal2025}}
\label{tab:stellar_parameters}
\begin{tabular}{ccccc}
\toprule
Parameter          & N24                      & RM21      & RM25                & This work                 \\ \midrule
$T_{eff, \text{WD}}/K$    & $11,000^{+1750}_{-1000}$  & $7500^{+5000}_{-2000}$    & 
$15,000^{+3750}_{-2000}$  &
$15,000^{+3812}_{-2750}$  \\

$T_{eff, \text{MS}}/K$    & $3800^{+800}_{-300}$      & $2800^{+200}_{-200}$      & 
$3200^{+100}_{-300}$      & $2800^{+200}_{-100}$      \\

$M_{\text{WD}}/M_{\odot}$ & $0.8^{+0.3}_{-0.3}$       & $0.4^{+0.1}_{-0.2}$       & 
$0.50^{+0.05}_{-0.06}$    &
$0.42^{+0.09}_{-0.08}$    \\

$R_{\text{WD}}/R_{\odot}$ & $0.010^{+0.007}_{-0.004}$ & $0.016^{+0.004}_{-0.003}$ &
$0.016^{+0.003}_{-0.002}$ &
$0.017^{+0.004}_{-0.003}$ \\ \bottomrule
\end{tabular}
\end{footnotesize}
\end{center}
\vspace{-0.4cm}
\end{table}

Concerning the excess of ELM WDs in our sample, a similar overabundance of low-mass WDs in the Gaia sample has been observed and discussed in \citet{mansergasetal2021b, hallakounetal2024} and \citet{lietal2025}.  This coincidence is not surprising, as all of these studies, including this one, focused on unresolved WDMS binaries. Due to Gaia's high angular resolution and the proximity of these stars (90\% of the WDMS in our sample are within 500 pc of the Sun), there is a clear selection bias toward close orbital configurations. Thus, a scenario in which WDs lose mass through stable mass transfer or a CE phase is more plausible.

However, it is also important to note the differences between these works and the present paper. \citet{hallakounetal2024} searched for WD companions at $\sim 1$ au of orbital separation from their MS host stars using an astrometric method. \citet{lietal2025} found most of their binaries in the bridge between the WD and the MS loci, with expected orbital separations of around $\sim 40$ au. The authors therefore argued that stable mass transfer was the most likely mechanism, discarding the CE phase because close orbits are required for it to occur.

In contrast, as explained in Section \ref{subsec:initialsample}, we focused our research strictly on the WD locus, applying astrometric and photometric cuts that biased our sample toward much tighter orbital configurations. Furthermore, as shown in Section \ref{sec:eclipsing} and \ref{subsec:stellar_masses}, we found extremely short orbital periods ($P \approx 0.5$ days) and small semi-major axes ($a \approx 0.02$ au) in the WDMS candidates with available light curves. These results align more closely with a CE phase that shrunk the orbit \citep{nebotetal2011, mansergasetal2011, mansergasetal2021b}. As discussed in Section \ref{subsec:stellar_masses}, this process could also explain the excess of low-mass WDs.

Nonetheless, we would like to emphasize that firmly establishing the PCEB nature of the objects in our sample requires a much deeper understanding of their orbital configurations, something beyond the scope of this study. We plan to verify these findings through follow‐up observations using ground‐based telescopes in future work.

During the preparation of this manuscript, \citet{santosgarciaetal2025} published a comprehensive statistical study of the  unresolved WDMS sample within 100 pc using population synthesis simulations. In their paper, they found that the majority of unresolved WDMS binaries are located in the main sequence ($\sim 90\%$), and in the intermediate region between the main sequence and the WD region (hereafter, the intermediate WDMS region, $\sim 10\%$). In fact, depending on the observational cuts they only expect to find between five and eight WDMS unresolved binaries within 100 pc in the WD region.

\begin{figure}
    \centering
    \includegraphics[width=0.49\textwidth]{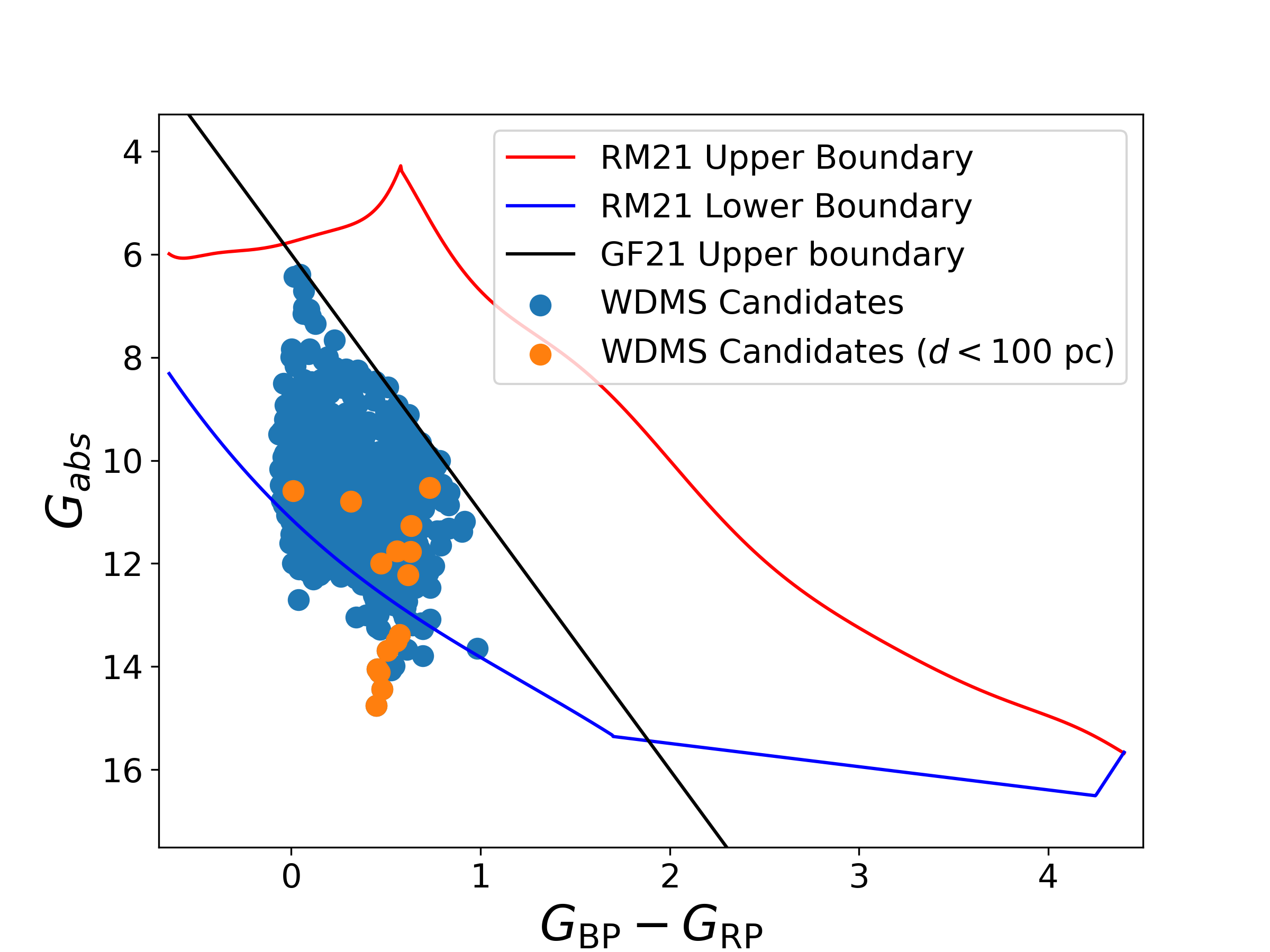}
    \caption{Gaia CMD with the 993 WDMS candidates found in this work, and the boundaries for the WD and WDMS region of \citet{gentilefusilloetal2021} and \citet{santosgarciaetal2025}.}
    \label{fig:wdms_region}
\end{figure}

To compare our results with their conclusions, we show in Figure \ref{fig:wdms_region} our 993 WDMS candidates plotted as blue dots in the Gaia CMD. A subset with the 15 sources that are within 100 pc are highlighted in orange. In black we show the upper boundary of the WD locus as defined in \citet{gentilefusilloetal2021} (hereafter GF21; see \eqref{WDlocus})) used in this work, and in red and blue the upper and lower boundaries of the intermediate WDMS region defined in \citet{mansergasetal2021b} and used in \citet{santosgarciaetal2025}. As can be seen, only seven WDMS candidates are located below their intermediate WDMS region, which is in excellent agreement with their study.

In summary, the WDMS candidates found in this work represent a new, different, and complementary population to that previously studied in the literature. However, follow-up observations of our candidates are necessary to confirm their binarity. If verified, our sample of 846 new WDMS binary candidates would increase the total number of known WDMS binary systems by $\sim 20\%$.

\section{Conclusions}

In this work, we have demonstrated the power of Self-Organizing Maps to unveil subtle regularities in the Gaia XP spectra. By combining dimensionality reduction and clustering, our SOM allowed us to identify a thousand of unresolved WDMS candidates in the \citet{gentilefusilloetal2021} catalog, of which 506 are new identifications.

The analysis presented here illustrates how the SOM can successfully separate WDMS binaries from single WDs based on spectral morphology. Even though our initial sample consists of WD companions that dominate the astrometry and photometry of their systems, our SOM demonstrates an excellent precision ($\sim 90\%$) in detecting red flux excesses. Unfortunately, the recall is very low because most of the MS companions in the WD locus are being outshined by its degenerated host, as we demonstrated here by comparing their luminosity ratios. Notwithstanding that, our recall it is sufficient to recover a third of the WDMS binaries present in the input sample.

We further validated 136 sources in our sample using the VOSA tool to fit binary SEDs with external UV, optical, and NIR photometry together with independent atmospheric models. As a result, we obtained a golden sample for which individual temperatures, luminosities, radii, and masses are estimated. 

Using these parameters, we characterized our sample as belonging to a population of atypical low-mass WDs that also include low-mass companions, primarily M dwarfs. A comparison with state-of-the-art WDMS catalogs shown that our method identified a complementary and previously undetected sample of WDMS binaries. This highlights the potential of Gaia DR3 (and the forthcoming DR4) XP spectra combined with unsupervised learning techniques to expand the known WDMS population.

Finally, a cross-match with the Gaia DR3 eclipsing binary catalog shows that at least 13 of our candidates have periodic variability, further supporting their classification as short-period interacting binaries with separations of the order of $\sim 0.01$ au. This subset of systems represents promising targets for follow-up studies.

\begin{acknowledgements}
We warmly thank the anonymous referee whose insightful
comments have greatly improved this paper. Scientific progress thrives on discussion and collaboration, and this paper is no exception. We sincerely thank the comments of our colleagues, Alberto Rebassa-Mansergas, Santiago Torres, Raquel Murillo-Ojeda, and Alejandro Santos-García during the 3rd meeting of the Iberian White Dwarf Workshop held in A Coruña in January 2025. However, any error is the sole responsibility of the authors. This work has made use of data from the European Space Agency (ESA) Gaia mission and processed by the Gaia Data Processing and Analysis Consortium (DPAC). Funding for the DPAC has been provided by national institutions, in particular, the institutions participating in the Gaia Multilateral Agreement. This work has made use of the Python package GaiaXPy, developed and maintained by members of the Gaia Data Processing and Analysis Consortium (DPAC) and in particular, Coordination Unit 5 (CU5), and the Data Processing Centre located at the Institute of Astronomy, Cambridge, UK (DPCI). This publication makes use of VOSA, developed under the Spanish Virtual Observatory (https://svo.cab.inta-csic.es) project funded by MCIN/AEI/10.13039/501100011033/ through grant PID2020-112949GB-I00.
VOSA has been partially updated by using funding from the European Union's Horizon 2020 Research and Innovation Programme, under Grant Agreement nº 776403 (EXOPLANETS-A). This research was funded by the Horizon Europe [HORIZON-CL4-2023-SPACE-01-71] SPACIOUS project, Grant Agreement no. 101135205, the Spanish Ministry of Science MCIN / AEI / 10.13039 / 501100011033, and the European Union FEDER through the coordinated grant PID2021-122842OB-C22. We also acknowledge support from the Xunta de Galicia and the European Union (FEDER Galicia 2021-2027 Program) Ref. ED431B 2024/21, ED431B 2024/02, and CITIC ED431G 2023/01. X.P. acknowledges financial support from the Spanish National Programme for the Promotion of Talent and its Employability grant PRE2022-104959 co-funded by the European Social Fund and E.V acknowledges funding from Spanish Ministry project  PID2021-127289NB-100 is also acknowledged.
\end{acknowledgements}

% \bibliography{sample631}{}

\clearpage
\begin{thebibliography}{99}

\begin{small}

\bibitem[Alam et al.(2015)]{alametal2015} Alam, S., Albareti, F.~D., Allende Prieto, C., et al.\ 2015, \apjs, 219, 12. doi:10.1088/0067-0049/219/1/12

\bibitem[\'Alvarez et al.(2022)]{alvarezetal2022} \'Alvarez, M.A., Dafonte, C., Manteiga, M. et al. 2022, Neural Comput \& Applic, 34, 1993–2006

\bibitem[Andrae et al.(2023)]{andraeetal2023} Andrae, R., Fouesneau, M., Sordo, R., et al. 2023, A\&A, 674, id.A27, 22.
\bibitem[Astropy Collaboration et al.(2022)]{astropy} Astropy Collaboration, Price-Whelan, A.~M., Lim, P.~L., et al.\ 2022, \apj, 935, 167. doi:10.3847/1538-4357/ac7c74

% {\bibitem[Badenas-Agusti et al.(2024)]{badenasagustietal2024} Badenas-Agusti, M., Vanderburg, A., Blouin, S., et al.\ 2024, \mnras, 527, 4515. doi:10.1093/mnras/stad3362}

\bibitem[Bailer-Jones et al.(2021)]{bailerjonesetal2021} Bailer-Jones, C.~A.~L., Rybizki, J., Fouesneau, M., et al.\ 2021, VizieR Online Data Catalog, 1352. I/352

\bibitem[Baraffe et al.(2015)]{baraffeetal2015} Baraffe, I., Homeier, D., Allard, F., et al.\ 2015, \aap, 577, A42. doi:10.1051/0004-6361/201425481

\bibitem[Bayo et al.(2008)]{bayoetal2008} Bayo, A., Rodrigo, C., Barrado Y Navascu{\'e}s, D., et al.\ 2008, \aap, 492, 277. doi:10.1051/0004-6361:200810395

\bibitem[B{\'e}dard et al.(2020)]{bedardetal2020} B{\'e}dard, A., Bergeron, P., Brassard, P., et al.\ 2020, \apj, 901, 93. doi:10.3847/1538-4357/abafbe

\bibitem[Belokurov et al.(2020)]{belokurovetal2020} Belokurov, V., Penoyre, Z., Oh, S., et al.\ 2020, \mnras, 496, 1922. doi:10.1093/mnras/staa1522

\bibitem[Benitez et al.(2014)]{benitezetal2014} Benitez, N., Dupke, R., Moles, M., et al.\ 2014, arXiv:1403.5237. doi:10.48550/arXiv.1403.5237

% \bibitem[Beuermann et al.(2013)]{beuermannetal2013} Beuermann, K., Dreizler, S., Hessman, F.~V., et al.\ 2013, \aap, 558, A96. doi:10.1051/0004-6361/201322241

\bibitem[Bianchi et al.(2017)]{bianchietal2017} Bianchi, L., Shiao, B., \& Thilker, D.\ 2017, \apjs, 230, 24. doi:10.3847/1538-4365/aa7053

\bibitem[Brinkworth et al.(2012)]{brinkworthetal2012} Brinkworth, C.~S., G{\"a}nsicke, B.~T., Girven, J.~M., et al.\ 2012, \apj, 750, 86. doi:10.1088/0004-637X/750/1/86

% \bibitem[Burrows et al.(2001)]{burrowsetal2001} Burrows, A., Hubbard, W.~B., Lunine, J.~I., et al.\ 2001, Reviews of Modern Physics, 73, 719. doi:10.1103/RevModPhys.73.719

% \bibitem[Carvalho et al.(2016)]{carvalhoetal2016} Carvalho, A., Marinho Jr., R. M., Malheiro, M., et al 2016, J. Phys.: Conf. Ser. 706 052016

\bibitem[Carrasco and Brunner(2014))]{carrascoandbrunner2014} Carrasco, K. M., Brunner, R. J. 2014, MNRAS, 438(4), 3409–3421

% \bibitem[Carrasco et al.(2021)]{carrascoetal2021} Carrasco, J. M., Weiler, M., Jordi, C., et al. 2021, A\&A, 652, A86

% \bibitem[Casewell et al.(2012)]{casewelletal2012} Casewell, S.~L., Burleigh, M.~R., Wynn, G.~A., et al.\ 2012, \apjl, 759, L34. doi:10.1088/2041-8205/759/2/L34

% \bibitem[Casewell et al.(2018)]{casewelletal2018} Casewell, S.~L., Braker, I.~P., Parsons, S.~G., et al.\ 2018, \mnras, 476, 1405. doi:10.1093/mnras/sty245

% \bibitem[Casewell et al.(2020)]{casewelletal2020} Casewell, S.~L., Belardi, C., Parsons, S.~G., et al.\ 2020, \mnras, 497, 3571. doi:10.1093/mnras/staa1608

% \bibitem[Chayer et al.(1995)]{chayeretal1995} Chayer, P., Fontaine, G., \& Wesemael, F.\ 1995, \apjs, 99, 189. doi:10.1086/192184

% \bibitem[Culpan et al.(2022)]{culpanetal2022} Culpan, R., Geier, S., Reindl, N., et al.\ 2022, \aap, 662, A40. doi:10.1051/0004-6361/202243337

\bibitem[Dafonte et al.(2018)]{dafonteetal2018} Dafonte, C., Garabato, D., Álvarez, M.A., Manteiga, M. 2018, Sensors, 18, 1419

\bibitem[De Angeli et al.(2023)]{deangelietal2023} De Angeli, F., Weiler, M., Montegriffo, P., et al. 2023, A\&A, 674, A2

% \bibitem[Delchambre et al.(2023)]{delchambreetal2023} Delchambre, L., Bailer-Jones, C.~A.~L., Bellas-Velidis, I., et al.\ 2023, \aap, 674, A31. doi:10.1051/0004-6361/202243423

\bibitem[Dufour et al.(2016)]{dufouretal2016} Dufour, P., Blouin, S., et al. 2016, arXiv:1610.00986 [astro-ph.SR]

\bibitem[Echeverry et al.(2022)]{echeverryetal2022} Echeverry, D., Torres, S., Rebassa-Mansergas, A., et al. 2022, A\&A, 667, A144

\bibitem[Epchtein et al.(1994)]{epchteinetal1994} Epchtein, N., de Batz, B., Copet, E., et al.\ 1994, \apss, 217, 3. doi:10.1007/BF00990013

\bibitem[Farihi et al.(2012)]{farihietal2012} Farihi, J., G{\"a}nsicke, B.~T., Steele, P.~R., et al.\ 2012, \mnras, 421, 1635. doi:10.1111/j.1365-2966.2012.20421.x

% \bibitem[Farihi et al.(2010)]{farihietal2010} Farihi, J., Barstow, M. A., Redfield, S., et al. 2010, MNRAS, 404, 2123
% \bibitem[Farihi \& Christopher(2004)]{farihiandchristopher2004} Farihi, J. \& Christopher, M.\ 2004, \aj, 128, 1868. doi:10.1086/423919

% \bibitem[Farihi et al.(2017)]{farihietal2017} Farihi, J., Parsons, S.~G., \& G{\"a}nsicke, B.~T.\ 2017, Nature Astronomy, 1, 0032. doi:10.1038/s41550-016-0032

% \bibitem[French et al.(2024)]{frenchetal2024} French, J.~R., Casewell, S.~L., Amaro, R.~C., et al.\ 2024, \mnras, 534, 2244. doi:10.1093/mnras/stae2121

\bibitem[Fustes et al.(2013a)]{fustesetal2013a} Fustes, D., Dafonte, C., Arcay, B. et al. 2013, Expert Syst Appl, 40(5), 1530–1541.

\bibitem[Fustes et al.(2013b)]{fustesetal2013b} 
Fustes, D., Manteiga, M., Dafonte, C. et al. 2013, A\&A, A7, 10.

\bibitem[Gaia Collaboration(2023)]{gaiacollaboration2023} Gaia Collaboration: Vallenari, A., Brown, A.G.A., Prusti, T., et al. 2023, A\&A 674, A1
% \bibitem[Garabato(2020)]{garabato2020} Garabato, D. 2020, PhD thesis. University of A Coruña

\bibitem[García-Zamora et al.(2023)]{garciazamoraetal2023} 
Garcia-Zamora   E. M., Torres   S., Rebassa-Mansergas   A., 2023, A\&A, 679, A127 

\bibitem[Garc{\'\i}a Zamora et al.(2025)]{2025arXiv250505560G} Garc{\'\i}a Zamora, E.~M., Torres Gil, S., Rebassa Mansergas, A., et al.\ 2025, arXiv:2505.05560. doi:10.48550/arXiv.2505.05560


\bibitem[Geach(2012)]{geach2012} Geach,. J. E. 2012, MNRAS, 419, 2633–2645

% \bibitem[Gentile-Fusillo et al.(2019)]{gentilefusillo2019}
% Gentile Fusillo, N.~P., Tremblay, P.-E., G{\"a}nsicke, B.~T., et al.\ 2019, \mnras, 482, 4570. doi:10.1093/mnras/sty3016

\bibitem[Gentile-Fusillo et al.(2021)]{gentilefusilloetal2021} 
Gentile Fusillo, N.~P., Tremblay, P.-E., Cukanovaite, E., et al.\ 2021, \mnras, 508, 3877. doi:10.1093/mnras/stab2672

\bibitem[Ginsburg et al.(2019)]{ginsburgetal2019} Ginsburg, A., Sip{\H{o}}cz, B.~M., Brasseur, C.~E., et al.\ 2019, \aj, 157, 98. doi:10.3847/1538-3881/aafc33

% \bibitem[Golovin et al.(2024)]{golovinetal2024} Golovin, A., Reffert, S., Vani, A., et al.\ 2024, \aap, 683, A33. doi:10.1051/0004-6361/202347767

\bibitem[Grondin et al.(2024)]{grondinetal2024} Grondin, S.~M., Drout, M.~R., Nordhaus, J., et al.\ 2024, \apj, 976, 102. doi:10.3847/1538-4357/ad7500

\bibitem[Hallakoun et al.(2024)]{hallakounetal2024} Hallakoun, N., Shahaf, S., Mazeh, T., et al.\ 2024, \apjl, 970, 1, L11. doi:10.3847/2041-8213/ad5e63

% \bibitem[Han et al.(2002)]{hanetal2002} Han, Z., Podsiadlowski, P., Maxted, P.~F.~L., et al.\ 2002, \mnras, 336, 449. doi:10.1046/j.1365-8711.2002.05752.x

% \bibitem[Han et al.(2003)]{hanetal2003} Han, Z., Podsiadlowski, P., Maxted, P.~F.~L., et al.\ 2003, \mnras, 341, 669. doi:10.1046/j.1365-8711.2003.06451.x

\bibitem[Hartmann et al.(2016)]{hartmannetal2016} Hartmann, S., Nagel, T., Rauch, T., et al.\ 2016, \aap, 593, A67. doi:10.1051/0004-6361/201628403

% \bibitem[Heber(2009)]{heber2009} Heber, U.\ 2009, \araa, 47, 211. doi:10.1146/annurev-astro-082708-101836

% \bibitem[Heber(2016)]{heber2016} Heber, U.\ 2016, \pasp, 128, 082001. doi:10.1088/1538-3873/128/966/082001

% \bibitem[Henden et al.(2015)]{hendenetal2015} Henden, A.~A., Levine, S., Terrell, D., et al.\ 2015, \aas, 225, 336.16

\bibitem[Iben et al.(1997)]{ibenetal1997} Iben I.J., Ritossa C., Garcia-Berro E., 1997, ApJ, 489, 772

% \bibitem[Izquierdo et al.(2020)]{izquierdoetal2020} Izquierdo, P., Toloza, O, Gänsicke, B. T., et al. 2020, MNRAS, 501(3), 4276-4288

% \bibitem[Jim\'enez-Esteban et al.(2023)]{jimenezestebanetal2023} Jiménez-Esteban, F. M., Torres, S., Rebassa-Mansergas, A., et al. 2023, MNRAS, 518, 5106.

\bibitem[Kao et al.(2024)]{kaoetal2024} Kao, M.~L., Hawkins, K., Rogers, L.~K., et al.\ 2024, \apj, 970, 181. doi:10.3847/1538-4357/ad5d6e

\bibitem[Kirkpatrick et al.(2021)]{kirkpatricketal2021} Kirkpatrick, J.~D., Gelino, C.~R., Faherty, J.~K., et al.\ 2021, \apjs, 253, 7. doi:10.3847/1538-4365/abd107

\bibitem[Kiviluoto(1996)]{kiviluoto1996} 
Kiviluoto, K. 1996, in Proc. Int. Conf. Neural Netw. (ICNN'96), Vol. 1, 294

% \bibitem[Kawka et al.(2015)]{kawkaetal2015} Kawka, A., Vennes, S., O'Toole, S., et al.\ 2015, \mnras, 450, 3514. doi:10.1093/mnras/stv821

% \bibitem[Klein et al.(2021)]{kleinetal2021} Klein, B. L., Doyle, A. E., Zuckerman, B., et al. 2021, ApJ, 914(1), id.61, 17

% \bibitem[Koester \& Wilken(2006)]{koesteretal2006} Koester, D., Wilken, D. 2006, A\&A, 453, 1051.

% \bibitem[Koester(2009)]{koester2009} Koester, D. 2009, A\&A, 498(2), 517-525.

\bibitem[Koester(2010)]{koester2010} Koester, D.\ 2010, \memsai, 81, 921

\bibitem[Kohonen(1982)]{kohonen1982}  Kohonen, T. 1982, Biol. Cybern. 43, 59-69

% \bibitem[Kramer et al.(2020)]{krameretal2020} Kramer, M., Schneider, F.~R.~N., Ohlmann, S.~T., et al.\ 2020, \aap, 642, A97. doi:10.1051/0004-6361/202038702

% \bibitem[Kuiper(1939)]{kuiper1939} Kuiper, G.~P.\ 1939, \apj, 89, 548. doi:10.1086/144075

\bibitem[Leibundgut \& Sullivan(2018)]{leibundgutandsullivan2018} Leibundgut, B. \& Sullivan, M.\ 2018, \ssr, 214, 57. doi:10.1007/s11214-018-0491-8

% \bibitem[Lema{{\^i}}tre et al.(2017)]{lemaitre2017} Lema{{\^i}}tre, G., Nogueira, F., Aridas, C. K. 2017, JMLR, 18(17), 1-5.

\bibitem[Li et al.(2025)]{lietal2025} Li, J., Ting, Y.-S., Rix, H.-W., et al.\ 2025, arXiv:2501.14494. doi:10.48550/arXiv.2501.14494

\bibitem[Lindegren et al.(2018)]{lindegrenetal2018} Lindegren, L., Hern\'andez, J., Bombrun, A., et al. 2018. A\&A, 616, id.A2, 25

% \bibitem[Luo et al.(2016)]{luoetal2016} Luo, Y.-P., N{\'e}meth, P., Liu, C., et al.\ 2016, \apj, 818, 202. doi:10.3847/0004-637X/818/2/202

% \bibitem[Luo et al.(2024)]{luoetal2024} Luo, Y., N{\'e}meth, P., Wang, K., et al.\ 2024, \apjs, 271, 21. doi:10.3847/1538-4365/ad1ab2

\bibitem[Magnier et al.(2020)]{magnieretal2020} Magnier, E.~A., Schlafly, E.~F., Finkbeiner, D.~P., et al.\ 2020, \apjs, 251, 6. doi:10.3847/1538-4365/abb82a

% \bibitem[Maldonado et al.(2020)]{maldonadoetal2020} Maldonado, R. F., Villaver, E., Mustill, A. J., et al. MNRAS, 499(2), 1854-1869

% \bibitem[Maldonado et al.(2021)]{maldonadoetal2021} Maldonado, R. F., Villaver, E., Mustill, A. J., et al. 2021. MNRAS, 501(1), L43-L48

\bibitem[Marocco et al.(2021)]{maroccoetal2021} Marocco, F., Eisenhardt, P.~R.~M., Fowler, J.~W., et al.\ 2021, \apjs, 253, 8. doi:10.3847/1538-4365/abd805

% \bibitem[Maxted et al.(2006)]{maxtedetal2006} Maxted, P.~F.~L., Napiwotzki, R., Dobbie, P.~D., et al.\ 2006, \nat, 442, 543. doi:10.1038/nature04987

\bibitem[Melis et al.(2012)]{melisetal2012} Melis, C., Dufour, P., Farihi, J., et al.\ 2012, \apjl, 751, L4. doi:10.1088/2041-8205/751/1/L4

\bibitem[Moe \& Di Stefano(2017)]{moeanddistefano2017} Moe, M. \& Di Stefano, R.\ 2017, \apjs, 230, 15. doi:10.3847/1538-4365/aa6fb6

\bibitem[Montegriffo et al.(2023)]{montegriffoetal2023} Montegriffo, P., Bellazzini, M., De Angeli, F., et al.\ 2023, \aap, 674, A33. doi:10.1051/0004-6361/202243709

\bibitem[Mowlavi et al.(2023)]{mowlavietal2023} Mowlavi, N., Holl, B., Lecoeur-Ta{\"\i}bi, I., et al.\ 2023, \aap, 674, A16. doi:10.1051/0004-6361/202245330

% \bibitem[Mustill et al.(2018)]{mustilletal2018} Mustill, A. J., Villaver, E., Veras, D., et al. 2018, MNRAS, 476(3), 3939-3955

\bibitem[Naim et al.(2009)]{naimetal2009} Naim, A., Ratnatunga, U., Griffiths, E. 2009, ApJ Suppl Series 111, 357

\bibitem[Nayak et al.(2024)]{nayaketal2024} Nayak, P.~K., Ganguly, A., \& Chatterjee, S.\ 2024, \mnras, 527, 6100. doi:10.1093/mnras/stad3580

\bibitem[Nebot G{\'o}mez-Mor{\'a}n et al.(2011)]{nebotetal2011} Nebot G{\'o}mez-Mor{\'a}n, A., G{\"a}nsicke, B.~T., Schreiber, M.~R., et al.\ 2011, \aap, 536, A43. doi:10.1051/0004-6361/201117514


\bibitem[Ordoñez-Blanco et al.(2010)]{ordonezetal2010} Ordoñe-Blanco, D., Arcay, B., Dafonte, C. et al. 2010, Lect Notes Essays Astrophys, 4, 97–102

% \bibitem[Pallas-Quintela et al.(2023)]{pallasquintela2023} Pallas-Quintela, L., Garabato, D., Manteiga, M., Dafonte, C. 2023, Highlights of Spanish Astrophysics XI, Proceedings of the XV Scientific Meeting of the Spanish Astronomical Society held on September 4--9, 2022, in La Laguna, Spain. M. Manteiga, L. Bellot, P. Benavidez, A. de Lorenzo-Cáceres, M. A. Fuente, M. J. Martínez, M. Vázquez Acosta, C. Dafonte (eds.)

\bibitem[Parsons et al.(2013)]{parsonsetal2013} Parsons, S.~G., G{\"a}nsicke, B.~T., Marsh, T.~R., et al.\ 2013, \mnras, 429, 256. doi:10.1093/mnras/sts332

% \bibitem[Parsons et al.(2017)]{parsonsetal2017} Parsons, S.~G., Hermes, J.~J., Marsh, T.~R., et al.\ 2017, \mnras, 471, 976. doi:10.1093/mnras/stx1610

\bibitem[Pecaut \& Mamajek(2013)]{pecautandmamajek2013} Pecaut, M.~J. \& Mamajek, E.~E.\ 2013, \apjs, 208, 9. doi:10.1088/0067-0049/208/1/9

% \bibitem[Pelletier et al.(1986)]{pelletieretal1986} Pelletier, C., Fontaine, G., Wesemael, F., Michaud, G., Wegner, G. 1986, ApJ, 307, 242

% \bibitem[Pelisoli et al.(2020)]{pelisolietal2020} Pelisoli, I., Vos, J., Geier, S., et al.\ 2020, \aap, 642, A180. doi:10.1051/0004-6361/202038473

\bibitem[P{\'e}rez-Couto et al.(2024)]{perezcoutoetal2024} P{\'e}rez-Couto, X., Pallas-Quintela, L., Manteiga, M., et al.\ 2024, \apj, 977, 31. doi:10.3847/1538-4357/ad88f5

% \bibitem[Phillips et al.(2020)]{phillipsetal2020} Phillips, M.~W., Tremblin, P., Baraffe, I., et al.\ 2020, \aap, 637, A38. doi:10.1051/0004-6361/201937381

\bibitem[Raddi et al.(2025)]{raddietal2025} Raddi, R., Rebassa-Mansergas, A., Torres, S., et al.\ 2025, arXiv:2502.01285. doi:10.48550/arXiv.2502.01285

\bibitem[Raghavan et al.(2010)]{raghavanetal2010} Raghavan, D., McAlister, H.~A., Henry, T.~J., et al.\ 2010, \apjs, 190, 1. doi:10.1088/0067-0049/190/1/1

\bibitem[Rebassa-Mansergas et al.(2010)]{mansergasetal2010} Rebassa-Mansergas, A., G{\"a}nsicke, B.~T., Schreiber, M.~R., et al.\ 2010, \mnras, 402, 620. doi:10.1111/j.1365-2966.2009.15915.x

\bibitem[Rebassa-Mansergas et al.(2011)]{mansergasetal2011} Rebassa-Mansergas, A., Nebot G{\'o}mez-Mor{\'a}n, A., Schreiber, M.~R., et al.\ 2011, \mnras, 413, 1121. doi:10.1111/j.1365-2966.2011.18200.x

\bibitem[Rebassa-Mansergas et al.(2012)]{mansergasetal2012} Rebassa-Mansergas, A., Nebot G{\'o}mez-Mor{\'a}n, A., Schreiber, M.~R., et al.\ 2012, \mnras, 419, 806. doi:10.1111/j.1365-2966.2011.19923.x

\bibitem[Rebassa-Mansergas et al.(2013)]{mansergasetal2013} Rebassa-Mansergas, A., Agurto-Gangas, C., Schreiber, M.~R., et al.\ 2013, \mnras, 433, 3398. doi:10.1093/mnras/stt974

\bibitem[Rebassa-Mansergas et al.(2016)]{mansergasetal2016} Rebassa-Mansergas, A., Ren, J.~J., Parsons, S.~G., et al.\ 2016, \mnras, 458, 3808. doi:10.1093/mnras/stw554

\bibitem[Rebassa-Mansergas et al.(2019)]{mansergasetal2019} Rebassa-Mansergas, A., Solano, E., Xu, S., et al.\ 2019, \mnras, 489, 3990. doi:10.1093/mnras/stz2423

\bibitem[Rebassa-Mansergas et al.(2021a)]{mansergasetal2021a} Rebassa-Mansergas, A., Maldonado, J., Raddi, R., et al.\ 2021, \mnras, 505, 3165. doi:10.1093/mnras/stab1559

\bibitem[Rebassa-Mansergas et al.(2021b)]{mansergasetal2021b} Rebassa-Mansergas, A., Solano, E., Jim{\'e}nez-Esteban, F.~M., et al.\ 2021, \mnras, 506, 5201. doi:10.1093/mnras/stab2039

\bibitem[Rebassa-Mansergas et al.(2025)]{mansergasetal2025} Rebassa-Mansergas, A., Solano, E., Brown, A.~J., et al.\ 2025, arXiv:2505.15895. doi:10.48550/arXiv.2505.15895

\bibitem[Ren et al.(2014)]{renetal2014} Ren, J.~J., Rebassa-Mansergas, A., Luo, A.~L., et al.\ 2014, \aap, 570, A107. doi:10.1051/0004-6361/201423689

\bibitem[Ren et al.(2018)]{renetal2018} Ren, J.-J., Rebassa-Mansergas, A., Parsons, S.~G., et al.\ 2018, \mnras, 477, 4641. doi:10.1093/mnras/sty805

\bibitem[Riello et al.(2021)]{rielloetal2021} Riello et al. 2021, A\&A, 649, id.A3, 33 pp.

\bibitem[Rogers et al.(2024)]{rogersetal2024} Rogers, L.~K., Bonsor, A., Xu, S., et al.\ 2024, \mnras, 527, 6038. doi:10.1093/mnras/stad3557

% {\bibitem[Ruz-Mieres(2024)]{ruzmieres2024} Ruz-Mieres, D. 2024, gaia-dpci/GaiaXPy: GaiaXPy v2.1.2, doi: 10.5281/zenodo.11617977}

\bibitem[Sana et al.(2014)]{sanaetal2014} Sana, H., Le Bouquin, J.-B., Lacour, S., et al.\ 2014, \apjs, 215, 15. doi:10.1088/0067-0049/215/1/15

\bibitem[Santos-Garc{\'\i}a et al.(2025)]{santosgarciaetal2025} Santos-Garc{\'\i}a, A., Torres, S., Rebassa-Mansergas, A., et al.\ 2025, \aap,  695, A161. doi:10.1051/0004-6361/202452989

% \bibitem[Schaffenroth et al.(2018)]{schaffenrothetal2018} Schaffenroth, V., Geier, S., Heber, U., et al.\ 2018, \aap, 614, A77. doi:10.1051/0004-6361/201629789

% \bibitem[Sion et al.(1983)]{sionetal1983} Sion, E. M., Greenstein, J. L., Landstreet, J. D., et al. 1983, ApJ, 269, 253

\bibitem[Skrutskie et al.(2006)]{skrutskieetal2006} Skrutskie, M.~F., Cutri, R.~M., Stiening, R., et al.\ 2006, \aj, 131, 1163. doi:10.1086/498708

% \bibitem[Steele et al.(2011)]{steeleetal2011} Steele, P.~R., Burleigh, M.~R., Dobbie, P.~D., et al.\ 2011, \mnras, 416, 2768. doi:10.1111/j.1365-2966.2011.19225.x

% \bibitem[Steele et al.(2013)]{steeleetal2013} Steele, P.~R., Saglia, R.~P., Burleigh, M.~R., et al.\ 2013, \mnras, 429, 3492. doi:10.1093/mnras/sts620

\bibitem[Sun et al.(2021)]{sunetal2021} Sun, Y., Cheng, Z., Ye, S., et al.\ 2021, \apjs, 257, 65. doi:10.3847/1538-4365/ac283a

% \bibitem[Swan et al.(2023)]{swanetal2023} Swan, A., Farihi, J., Su, K. L-U., Desch, S. J. 2024, MNRAS: letters, 529(1), L41–L46

\bibitem[Swan et al.(2024)]{swanetal2024} Swan, A., Farihi, J., Su, K.~Y.~L., et al.\ 2024, \mnras, 529, L41. doi:10.1093/mnrasl/slad198


\bibitem[Torres et al.(1998)]{torresetal1998} Torres, S., García-Becerro, E., Isern, J. 1998, ApJ, 508, L71

% \bibitem[Trierweiler et al.(2023)]{trierweileretal2023} Trierweiler, . L., Doyle, A. E., Young, E. D. 2023, PSJ, 4(8), id.136, 13

% \bibitem[van Roestel et al.(2021)]{vanroesteletal2021} van Roestel, J., Kupfer, T., Bell, K.~J., et al.\ 2021, \apjl, 919, L26. doi:10.3847/2041-8213/ac22b7

% \bibitem[Veras et al.(2024)]{verasetal2024} Veras, D., Mustill, A., Bonsor, A. 2024,  
% Reviews in Mineralogy and Geochemistry, 90(1), 141-170

\bibitem[Vettigli(2018)]{vettigli2018} Vettigli, G. 2018, \url{https://github.com/JustGlowing/minisom/}

% \bibitem[Vincent et al.(2023)]{vincentetal2023} Vincent, O., Bergeron, P., Dufour, P. 2023, MNRAS, 521, 760

\bibitem[Vincent et al.(2024)]{vincentetal2024} Vincent, O., Barstow, M. A., Jordan, S., et al. 2024, A\&A, 682, A5

% \bibitem[Viscasillas V{\'a}zquez et al.(2024)]{viscasillasetal2024} Viscasillas V{\'a}zquez, C., Solano, E., Ulla, A., et al.\ 2024, \aap, 691, A223. doi:10.1051/0004-6361/202451247

\bibitem[Wang \& Han(2012)]{wangandhang2012} Wang, B. \& Han, Z.\ 2012, \nar, 56, 122. doi:10.1016/j.newar.2012.04.001

\bibitem[Way and Klose(2012)]{wayandklose2012} Way, M., Klose, C. 2012, PASP, 124

% \bibitem[Weiler et al.(2023)]{weileretal2023} Weiler, M., Carrasco, J. M., Fabricius, C., Jordi, C. 2023, A\&A 671, A52

\bibitem[Willems \& Kolb(2004)]{willemsandkolb2004} Willems, B. \& Kolb, U.\ 2004, \aap, 419, 1057. doi:10.1051/0004-6361:20040085

\bibitem[Wilson et al.(2019)]{wilsonetal2019} Wilson, T.~G., Farihi, J., G{\"a}nsicke, B.~T., et al.\ 2019, \mnras, 487, 133. doi:10.1093/mnras/stz1050

\bibitem[Winters et al.(2019)]{wintersetal2019} Winters, J.~G., Henry, T.~J., Jao, W.-C., et al.\ 2019, \aj, 157, 216. doi:10.3847/1538-3881/ab05dc

% \bibitem[Wright et al.(2010)]{wrightetal2010} Wright, E.~L., Eisenhardt, P.~R.~M., Mainzer, A.~K., et al.\ 2010, \aj, 140, 1868. doi:10.1088/0004-6256/140/6/1868

% \bibitem[Xu et al.(2024)]{xuetal2024} Xu, S., Rogers, L., Blouin, S. 2024, arXiv:2404.15425

\bibitem[Xu \& Jura(2012)]{xuandjura2012} Xu, S. \& Jura, M.\ 2012, \apj, 745, 88. doi:10.1088/0004-637X/745/1/88

\bibitem[Zhao et al.(2012)]{zhaoetal2012} Zhao, J.~K., Oswalt, T.~D., Willson, L.~A., et al.\ 2012, \apj, 746, 144. doi:10.1088/0004-637X/746/2/144

% \bibitem[Zuckerman et al.(2007)]{zuckermanetal2007} Zuckerman, B., Koester, D., Melis, C., et al. 2007, ApJ, 671(1), 872-877

\end{small}

\end{thebibliography}
\bibliographystyle{aasjournal}

\end{document}